# Description and Recognition of Regular and Distorted Secondary Structures in Proteins Using the Automated Protein Structure Analysis Method

Sushilee Ranganathan [1], Dmitry Izotov [1], Elfi Kraka [1], and Dieter Cremer *[1,2]

*[1]Department of Chemistry, University of the Pacific, 3601 Pacific Avenue, Stockton, CA 95211, USA, [2] Department of Physics, University of the Pacific, 3601 Pacific Avenue, Stockton, CA 95211, USA.*

*(E-mail: dcremer@pacific.edu)*

**Abstract**: The Automated Protein Structure Analysis (APSA) method, which describes the protein backbone as a smooth line in 3-dimensional space and characterizes it by curvature $\kappa$ and torsion $\tau$ as a function of arc length *s,* was applied on 77 proteins to determine all secondary structural units via specific $\kappa(s)$ and $\tau(s)$ patterns. A total of 533 $\alpha$-helices and 644 $\beta$-strands were recognized by APSA, whereas DSSP gives 536 and 651 units, respectively. Kinks and distortions were quantified and the boundaries (entry and exit) of secondary structures were classified. Similarity between proteins can be easily quantified using APSA, as was demonstrated for the *roll* architecture of proteins ubiquitin and spinach ferridoxin. A twenty-by-twenty comparison of *all-alpha* domains showed that the curvature-torsion patterns generated by APSA provide an accurate and meaningful similarity measurement for secondary, super-secondary, and tertiary protein structure. APSA is shown to accurately reflect the conformation of the backbone effectively reducing 3-dimensional structure information to 2-dimensional representations that are easy to interpret and understand.

## 1. Introduction

A qualitative and quantitative understanding of protein structure is an essential requirement for unraveling the relationship between protein shape and protein functionality. Numerous investigations have been carried out for this purpose. [1-13] At the more qualitative level, the ribbon representations made popular by Richardson [1] have given a visual entry to protein structure. The task of bringing these representations from the qualitative to the quantitative level of understanding requires a tedious analysis of conformational features and their representation in 3-dimensional (3D) space in form of symbolic or mnemonic devices. Attempts in this way that describe a specific fold with prior knowledge of its shape and properties do not fulfill the objective of finding a general concept of protein structure directly. Among such investigations is one that describes viral capsid jellyroll topology as wedges [2] and another that obtains orientation angles for the TIM-barrel motif from 7 domains [3]. There are other studies that provide detailed accounts of the various types of arrangements of helices [4,5] and β-strands [5] in folds. Though these descriptions throw light on the folding and function of a specific set of proteins, they use different approaches and levels of simplification, preventing their use for automated analysis and classification of proteins in general.

Among those methods that do classify all proteins in the Protein Data Bank (PDB) [6] many are not fully automated, such as CATH [7] and SCOP [8] that require manual intervention for analysis and decision-making. Some of the fully automated methods use more than one criterion (for example, the secondary STRuctural IDEntification (STRIDE) method [9] uses $\varphi,\psi$ angles and hydrogen bonding) or arbitrary parameters (for example, the Dictionary of Secondary Structure of Proteins (DSSP) [10] works with arbitrary energy cut-offs that determine the presence of hydrogen bonds) as the basis of analysis. The analysis of the 3D position of individual backbone points such as the $C_\alpha$-atoms using distance masks (DEFINE [11]) or distance matrices (among other criteria) [12] implies that discrete sets of data points miss important features of protein structure. If the $\varphi$ and $\psi$ backbone dihedral angles [13] are used as discrete parameters, the nontrivial task emerges to translate a multitude of dihedral angles into a general conformational concept for the purpose of understanding protein structure and functionality. Again, this task has so far not been satisfactorily solved. Hence, a simple, fully automated method that accurately reflects the conformation of the entire polypeptide chain, is easy to interpret, and relates to the 3D shape of the protein is needed.

Recently, we presented a new method for the automated protein structure analysis (APSA) that is based on a 2-step approach of describing and categorizing conformational features of proteins. [14] i) The protein backbone is simplified to a smooth, continuous line in 3D-space. ii) The curving and twisting of the backbone line is quantified by the curvature and torsion functions κ(s) and τ(s) where the parameter *s* gives the arc length of the backbone line. The diagrams of κ(s) and τ(s) adopt typical patterns that make identification of protein secondary structural units easy. [14] In addition, they quantitatively identify all deviations and distortions from the ideal and provide an easy classification and identification of non-regular structural features. A curvature or torsion peak representing the conformation of a residue in a protein reflects also conformational features of the neighboring residues. This complies with the fact that it takes more than one residue (represented in APSA by an $C_\alpha$ atom) to determine local shapes such as α-helix and β-strand. APSA works on this principle. Therefore, in the κ(s) and τ(s) diagrams, an 'ideal helix peak' of a particular $C_\alpha$ atom reflects the ideal (or close-to-ideal) helix arrangement of the two neighboring $C_\alpha$ atoms as well thus constituting an *ideal conformational environment*.

A search for amino acids in ideal conformational environments showed that only 63 % of all residues in α-helices and 49 % in β-strands comply with this conformational criterion. This discrepancy between the total number of secondary structural units identified in proteins and the number of ideal helices and ß-strands is partly the reason for disparities that occur among the secondary structure assignments of several automated methods discussed in literature [15]. We also demonstrated how the extended and helical nature of turns is accurately described and identified with the help of their κ(s) and τ(s) diagrams. [14] Thus, APSA was shown to be a qualitative as well as quantitative tool for protein structure analysis that projects the 3D conformational features into 2D representations.

In this work, APSA is applied to a set of 77 natural proteins with the objective of quantitatively describing distortions and deviations of helices and ß-strands from their ideal conformations. This involves the analysis and categorization of helix caps, entry and exit points of secondary structural units, kinks, bends, and breaks on the basis of the κ(s) and τ(s) diagrams. In this connection, the speed of automation, the reliability of the secondary structure assignment, and APSA's versatility in describing varied backbone conformations from diverse proteins will be tested. Throughout the investigation APSA assignments will be compared with DSSP, [10] which is a widely accepted secondary structure assignment method. A single protein (ubiquitin) will be selected from the set of 77 proteins to demonstrate the application of APSA in detail with respect to the characterization of all secondary structure and turn residues. Similar features seen between ubiquitin and spinach ferridoxin from the κ(s) and τ(s) diagrams will guide the way for a simple and effective protein structure comparison based on the treatment of proteins in form of continuous conformational patterns rather than a set of discrete conformational parameter points.

## 2. Computational Procedures

As described in Ref. [14], APSA is based on the representation of the protein backbone in form of a regularly parameterized smooth curve in 3D space. For this purpose, a coarse-grained image of the backbone is constructed where each residue is represented by its $C_\alpha$-atom. These positions are used as anchor points in 3D-space and are connected by a cubic spline function. The cubic spline gives the simplest parameterization of the backbone (compared to higher spline functions); it is computationally robust and easy to implement. Using the methods of differential geometry, the backbone line is described by means of three scalar parameters, curvature κ, torsion τ, and the arc length *s*. The functions κ(s) and τ(s) are generated by APSA for each protein from its coordinates taken from the PDB. [6]

As shown in Ref [14], curvature and torsion values calculated from the spline are not sensitive to the uncertainties in the atomic coordinates as long as the refinement of the X-ray structural analysis is equal or smaller than 2 Å. The mathematical and physical aspects of the APSA protocol were found to reasonably represent the details of structure and also include global features such as chirality and orientation of structural units in 3D-space. For technical details relating to quantification of sensitivity and properties of the spline fit, we refer to Ref. [14].

A set of 77 proteins (78 chains) listed in the Supporting Information was selected from the PDB [6] including proteins from the four classes of the CATH classification system [7] i.e., '*mainly alpha*', '*mainly beta*', '*mixed alpha-beta*' and '*few secondary structures*'. Only X-ray structures having a refinement of 2.0 $Å^{-1}$ or less were selected. Proteins having breaks in the structure, missing amino acids or alternate locations for $C_\alpha$ atoms were avoided. The proteins used for the APSA description are of different sizes with differing lengths of helices, β-sheets, and loop regions. They have one or more domains on single or several chains and in addition, are monomer or parts of multimeric structure. In the final dataset, the *mainly alpha* class includes 26 different proteins (and 28 domains), the *mainly beta* class 24 other proteins (and 26 domains), and the *alpha and beta* class, 23 new proteins (and 30 domains). Two new proteins (and 3 domains) are included under the *few secondary structures* class for insights into any standard conformations assumed by these domains. Some popular architectures are represented by a greater number of domains, like the *orthogonal bundle*, though rare architectures such as the *box* under the *α and β class* are also considered. In addition, various ratios of helices to β-sheets are represented within each architecture. In some cases, sets of identical or very similar proteins are purposely included in the analysis for similarity comparisons and so are some proteins with distinct supersecondary motifs.

## 3. Results and Discussions

In Table 1, the average, minimum and maximum κ and τ values of 5 α-helices (leaving out the N-terminal and C-terminal residue) and 8 β-strands, all of them free of specific distortions, are recorded. All but one of the β-strands chosen have negative torsion values indicative of a left-handed torsion along the ß-strand. [14] The eighth ß-strand is an example for one with a right-handed torsion. The average κ values for the α-helices were determined from 10 equidistant points located along the protein backbone between 2 successive $C_\alpha$ atoms where the latter were included into the set of equidistant points (they are not simply the average of minimum and maximum value).

Tables 1 and 2, Scheme 1, Figure 1

Utilizing the values listed in Table 1, nine ranges of κ(s) and τ(s) values arranged in four 'windows' were set up to create rules for automated structure recognition (Table 2, Scheme 1). The values in Table 2 were found to strike the right balance by accounting for irregular boundaries of secondary structures without losing geometric details. For α-helices, the *working* values differed from the *ideal* values obtained from earlier evaluations [14], wherein the κ(s) and τ(s) ranged from 0.3 $Å^{-1}$ to 0.56 $Å^{-1}$ and 0.08 $Å^{-1}$ to 0.19 $Å^{-1}$ respectively (Figure 1a, Table 2). These ranges have been relaxed for natural helices such that κ(s) ranges from 0.23 $Å^{-1}$ to 0.67 $Å^{-1}$ and τ(s), from 0.05 $Å^{-1}$ to 0.24 $Å^{-1}$ (window 2, Table 2). For example, the body of the helix in 1U4G starting at leucine 135 (κ(s) and τ(s) diagrams of Figure 1b) shows deviations from ideal α helical values not sufficiently significant to be considered as a special case of distortion. (Slightly distorted helices are shown in Figures 1c to 1m.)

The κ and τ values of the first amino acid are different from those of the body of the helix (Figure 1b), which is considered by defining window 1 for the 'starter' residue (Table 2). The high values of window 1 reflect the fact that the backbone enters into the helix from a relatively straight region by veering sharply into it. Similarly, the $C_α$ atom of the last amino acid belonging to the helix exit (Scheme 2) is at the centre of the smooth transition from a curved helical segment into the relatively straight segment of the following backbone (Figure 1b). The first $C_α$ point may either lie towards the body of the helix, in which case it has a positive τ value, or may lie away from the helix axis, when it shows τ values changing from negative to positive (see also Section 3.2.4). Both cases lead to $τ_{max} < 0.4$ $Å^{-1}$ (Table 2). κ-Values are not included into window 1 because they are too unspecific to facilitate identification of the helix starter residue.

Scheme 2

Naturally occurring β-strands are mostly twisted or bent and seem to be influenced easily by the surrounding turns and structures. This is especially true in the case of the β sheet occurring in folds such as the *roll* or the *ß-barrel*. These effects are clearly reflected in the κ- and τ-pattern of naturally occurring ß-strands (windows 3 and 4). The κ(s) peak lengths are large (larger than those of a helix) thus yielding higher peaks (0.5 to 1.4; helices: $κ < 0.65$ $Å^{-1}$, Table 2) and a much lower base (see Scheme 1) with values close to zero (helices: $κ > 0.25$ $Å^{-1}$). For the purpose of distinguishing the curvature of ß-strands from that of helices, a split window is used (Scheme 1, Table 2). It is noteworthy that for the ideal left-handed ß-strand (Figure 1n), the curvature values are $\leq 1.0$ $Å^{-1}$ (Table 2).

The $\tau$(s) peaks of the β-strands are also recognized by their base and tip values tested by a split $\tau$-window (Tables 1 and 2) where one part accounts for the base with values close to zero and the other for the extremes. In addition, one has to consider the sign of the $\tau$(s)-peaks, which indicates a left-handed (- sign, troughs; window 3 in Scheme 1, Table 2) or right-handed strand (+ sign, peaks; window 4). The troughs of the negative $\tau$(s) have so low-lying minima that it is sufficient to give just an upper boundary of $\tau$ (-0.75 $Å^{-1}$, Table 2). The positive peaks of the right-handed ß-strands (window 4) have somewhat different $\kappa$-ranges (0.4 to 0.9 $Å^{-1}$) and a different $\tau$-window. The values of the peak bases are found between 0.001 and 0.15 $Å^{-1}$ and the peak maxima are > 0.75 $Å^{-1}$ (Table 2, Scheme 1).

Additional windows can be defined for left-handed α-helices, right-handed $3_{10}$-, and π-helices. However, due to the fact that these structures occur relatively seldom in the ideal conformations, we refrain at this stage from setting up suitable $\kappa$ and $\tau$ ranges. Instead, we operate with the curvature and torsion values obtained for the ideal structures described in our previous work. [14]

### 3.1 Application of the APSA Windows

A summary of all α helices and β strands found among the 77 proteins (78 chains) investigated (see also Section 2 and Supporting Information) is presented in Table 3. Special forms of secondary structural units are shown in Figures 1a to 1p, which contain the calculated $\kappa$(s) and $\tau$(s) diagrams and a VMD [16] representation of the 3D structure. The 77 proteins possess a total of 547 α-helices and 656 β-strands according to DSSP assignments [10] that are based on hydrogen bonding patterns. APSA identifies 543 helices and 654 ß-strands where the numbers do not necessarily indicate a close match with the DSSP assignments. As is detailed in Table 3, differences are found for 20 helices and 46 ß-strands, which leads to an agreement in 96 % and 93 % of all helical and ß-strand environments respectively.

Table 3

APSA, contrary to DSSP, clearly identifies all distorted shapes via their graphic patterns of $\kappa$ and $\tau$ values falling outside the strict windows of Table 2. For example, 20 split or kinked helices are recognized, though in these cases DSSP had labeled them as one continuous helix (Table 3). The distortion in $\tau$(s) at leucine 89 in 1V54 corresponds to a change in the helix orientation as is confirmed by the ribbon representation (Figure 1c). Four DSSP-labeled H-bonded turns appear α-helical in the κ-τ diagrams and so do other regions that have no secondary structures assigned. Twelve structures recognized as "α-helices" are found to deviate from the regular pattern and hence are commented as being "distorted" in Table 3, though they show overall helical shapes (see, e.g., Figure 1d). Structures identified as "$3_{10}$-helices" by DSSP do not possess a unique APSA pattern, which is in line with descriptions of variable $3_{10}$-helix geometries in peptides as given in the literature. [1], [17]

APSA finds 654 undistorted β-strands. These correspond to β-strands alone; 'isolated β-bridges' of DSSP are not included for reasons of simplicity (excluding the need to analyze loops and turns). 22 new β-strands are found (with start and end residue ranges slightly shifted along the backbone; Table 3) and in 22 cases, the strands are geometrically distorted due to different bending and twisting of amino acids not typical of ideal β-strands (see Figure 1o for a distorted ß-strand in 1RIE). In two situations, adjacent strands merged into one continuous strand resulting in the loss of 1 strand in each case.

### 3.2 The APSA Description of Helices and their Distortions

The κ(s) and τ(s) diagrams precisely reflect the range of distortions for the helices investigated. There are some regions of the backbone where the geometry, though close to an α-helix, is intermediate between the α and $3_{10}$ conformation. Such distortions can occur both in the body of helices and toward their ends. *Splits and kinks* in the body of helices have been extensively studied and accounted in literature. [18]-[20] The degree of kink can be quantified using the κ and τ values. An example is the helix between E80-G97 in the E chain of bovine heart cytochrome C oxidase (1V54) [6] (Figure 1c). The corresponding τ diagram identifies the amino acid (L 89) responsible for the kink. The height of the τ peak (just over 0.5 $A^{-1}$) indicates that the kink is still helical, but the κ and τ values are close to those of a $3_{10}$-helix. [14] A more drastic kink, as in cytochrome P450 (2CPP), produces a corresponding strong disturbance in both κ(s) and τ(s) (see Figure 1e).

*Helical distortions* can be considered as regions where the backbone is still helical, but does not belong to the well-defined conformations of the $3_{10}$-, α-, or π-helices. The τ(s) diagrams of these regions are interspersed with extended peaks indicating stretching of the helix. These distended helices regions have traces of overall helicity and the coiling of the entire backbone into a helix becomes visible only as a global characteristic.

The analysis of secondary structures in proteins is often confronted with the problem of ambiguous boundaries. Early investigation [1] have documented that the ends of helices are different from the body. For this reason, secondary structure assignment methods must treat the amino acids belonging to these 'cap'-like structures with some caution. For example, some dihedral angle-based methods [21] [22] analyzed helices by discarding amino acids that took up any set of values lying outside predefined regions of the Ramachandran plot. A detailed geometry-based analysis of such regions would throw more light on this problem and also suggest a systematic and uniform way of classifying and handling them in future. This is possible using calculated κ and τ values of these regions. In some cases, the difference between the body and the termini of a helix is so strong that it might be considered as a turn rather than as an extension of the helix whereas in other cases it may be very subtle (Figure 1f). Therefore, we will explicitly discuss this problem in 3.2.3 and 3.2.4.

***3.2.1 The $3_{10}$-helix conformation**:* $3_{10}$-helices were first reported in 1941 [23]. The N termini of $3_{10}$- and α-helices have been studied and compared [24] with specific amino acid propensities and preferences. The latter were related to functionality and probable progression of protein folding along the α-helical axis [25]. It has been proposed [1] that the occurrence of $3_{10}$-conformations at the ends of helices serve the purpose of tightening the α-helix from uncoiling and losing its orientation. It has also been documented that $3_{10}$-helices could smoothly uncoil into α-helices and vice versa because the corresponding Ramachandran regions are *allowed* for this transformation. [26] This observation suggests the possibility of functional importance to these regions. Thus amino acids in a $3_{10}$-cap, whether at the N or C terminus of the helix, fulfill the purpose of a tighter coiling and stabilizing the ends of an α-helix.

The $3_{10}$-helices occurring at the C termini of α-helices can have an $\alpha_\pi$-conformation (π-conformation mixed into an α-helix), with H-bonding resembling the α-helix pattern and the slightly tilted conformation resembling the π helices. [1] For example, the region 8-17 of myoglobin (5MBN) has such an $\alpha_\pi$-character, which is confirmed by the corresponding κ and τ patterns (Figure 1g). The difference between an α and a $3_{10}$ N-terminus is that in the former case τ reaches up to 0.4 $Å^{-1}$ whereas in the latter case, it ranges from 0.4 to 0.56 $Å^{-1}$, [14] thus reflecting the different rise per amino acid of both structures along the helix axis. From the κ(s) and τ(s) diagrams of APSA, a smooth transition is often seen from the α-helix through the $3_{10}$-helix into the extended regions of turns or β-strands. The α-helix (3.6 amino acids per turn) possesses an average curvature peak length of 0.56 – 0.3 = 0.26 $Å^{-1}$ (Table 2) and therefore is more relaxed than a $3_{10}$-helix (3 amino acids per turn) with an average κ peak length of 0.81 – 0.28 = 0.53 $Å^{-1}$. [14] The transition from the $3_{10}$-helix conformation into the β-strand can be understood on the basis that the well-extended β-strand can be viewed as a helix with 2-amino acids per turn thus leading to higher κ-peaks than those of a $3_{10}$-helix (up to 1.0 compared to 0.8 $Å^{-1}$ in the latter case; Table 2 and Ref [14]). This trend can be partly seen in 1QTE (Figure 1h) at the (positive) τ-peaks corresponding to amino acid methionine 28, leucine 32, and aspartate 34.

***3.2.2 The π-helix conformation***. Though it has been known over the years that π-helices are rare, there are conflicting results [27] that indicate their occurrence to be as frequent as one out of every 10 helices in the PDB [6]. It is also discussed how H-bonding and amino acid preferences can be used to characterize π -helices and enumerates important associated functionalities such as specific ligand binding. [27] Some studies [26] consider the π– and the $3_{10}$-helix as folding intermediates in the formation of the α-helix; the α- and the $3_{10}$-helices have been described to share a common initiation paths while folding. [25] In the set of 77 proteins investigated by APSA, a pure π-region was not observed although π -character was found to be mixed into some of the helix caps (see 3.2.1).

***3.2.3 Helix termini as described by APSA***. In literature [24] the term helical *cap* is used for the last helical amino acid, whereas helix *terminus* (and in other literature [1] the same term *cap*) is used to denote a few amino acids towards the end of the helix (see Scheme 2) indicating that they are not always sharply defined. It should be noted that the terms *cap*, *terminus*, and *end* are used interchangeably and thus become loosely defined in literature. In the APSA investigation, the terms become equivalent because the spline fitting ensures that the κ(s)- and τ(s)-functions at every amino acid reflect the conformation of the neighboring amino acids. Amidst all the discussion about the occurrence, distribution, property, and details of helices and helix caps, there is no systematic classification of these structures based on just the geometry. APSA considers caps as a special case of "distortions" occurring toward the termini of helices. From the κ(s) and τ(s) diagrams of various protein segments it becomes evident that the cap at the terminus conformationally spreads over neighboring amino acids in either direction, and can be identified using torsion τ(s) alone. Utilizing the APSA results, the termini can be broadly divided into three different types.

Scheme 2

*i) α-Terminus*: This is a segment of α helix broken off from its body. About 3 or 4 amino acids of the α-helix are cut off from the rest and oriented toward a direction different from that of the helix. α-Termini can show some standard distortions and resemble the α-helix only by average κ and τ values. They include the $\alpha_\pi$-type of structures (Figure 1g).

*ii) Tighter terminus*: Such a terminus has a larger κ value, thus including $3_{10}$-caps and the distortions that are narrower in diameter than an α-helix loop. Some caps of mixed geometry are distorted with only the bare remnants of helicity resembling a completely stretched spring. In these cases, defining the cap and differentiating it from a loop region becomes difficult. The κ(s)-τ(s) diagrams reflect the true state of the backbone in a graphical way that aids the analysis and recognition of complicated patterns. Some examples of tighter termini are presented in Figures 1i-1*l*.

*iii) Looser terminus*: This terminus is more relaxed with a larger α-helix diameter and therefore includes a typical π cap or related distortions. Figure 1d shows the ending of the helix in 1QOY (hemolysin E) with a looser terminus starting at leucine 24. The larger diameter of the terminus increases the flexibility of the backbone to some extent introducing alternating high and low τ(s) values typical of helical yet more planar curves. Looser termini and α-termini appear to occur much less frequently than tighter helix termini.

3.2.4 ***Helix entries and exits***. Some helix entries and exits have been described in literature based on ϕ-ψ values and amino acid properties [29]. By APSA, the $C_\alpha$ atom of the amino acid prior to the starting of the helix is considered to be the "entry" (Scheme 2). The polypeptide chain can enter into the helix in either a left- or right-handed fashion. The left-handed entry is found more frequently (Figure 1b, f , i-k) and is the point of chain reversal from strongly negative τ(s) (left-handed torsion) through τ(s) ~ -0.1 $Å^{-1}$ at $C_\alpha$ to positive τ(s) values (right-handed helix torsion). The right-handed entry (Figure 1c, h, m) leads to no chain reversal and therefore the τ(s) remains positive. They have large values for curvature peak heights (with low minima; Figure 1m) resembling the peaks of extended conformations, whereas the following κ-minima are relatively large and slightly helical giving the impression as if the helix has been stretched to increase its pitch (Figure 1m).

*Helix exits*, much like the entries, can have positive or negative torsion, where again the latter are more frequent (Figures 1f, 1m). The positive exit in Figures 1h and 1*l* continues in the same overall direction of the helix whereas the negative exit appears to "peel away" from the helical formation (see inset of Figure 1b).

### 3.3 The APSA Description of Extended Structures and their Distortions

In Section 3.1, we showed that the series of τ peaks representing the β-regions can be either positive or negative where the sign gives the overall orientation of the strand in 3D (left- or right-handed twist). On a more detailed note, a β-strand could be considered to have 'local' and 'global' twisting. The 'local' twist is given by the arrangement of $C_\alpha$ atoms along the strand and the 'global' twist refers to the twisting of the whole β-ribbon. Both local and global twisting of the strand contributes to the torsion, the former being dominant. The global twisting is relatively small and does not produce any noticeable impact on the overall torsion value. The sign of the strand itself is indicative of the direction it points in 3D with respect to the last point in the preceding structure (strand, turn, loop, helix).

Figure 1p shows three pairs of helices from different proteins and demonstrates the handedness of local twisting in ß-strands. The first pair from parvalbumin (1CDP, 1-17) has two helices separated by two β-troughs, the second pair from hemerythrin (1HMD, 55 to 77) by three, and the third pair from ribosomal protein (1CTF, 67 to 90) by four. For the addition of every β-trough, the second helix does not only undergo a translation, but also a rotation: the relative orientations of the helices reveals that the first would rotate into the second, which would rotate into the third in a left-handed fashion. An addition of one more β-trough in the turn region would point the second helix in the same direction as in 1CDP, hence indicating pattern repetition for the addition of every fourth β-trough. The positive β-peaks (not shown) were found to have the same effect in the opposite direction of rotation, confirming that extended regions have local twisting and are not flat ribbons. Application of APSA to extended regions reveals that they are separated by numerous one residue-long kinks that bend the strands by less than 90º, though these are sometimes considered as supersecondary structures. [29a] It is interesting to note that a range of τ-peaks can be obtained for all intermediate structures ranging from a planar 90° strand (τ(s) close to 0, large κ(s)) to a strand that is bent strongly out of plane of the ß-ribbon (close to the torsion of a $3_{10}$-helix). When looking end-on (along the axis of a helix or ß-strand), a helix looks like a circle and a ß-strand like an ellipsis (rather than just a straight line as is often shown in textbooks for reasons of simplification). The plane of the ß-ribbon refers to that defined by the strand axis and the major axes of the ellipsis.

**3.3.1** ***β-Strand entries and exits***: The positive entries into β-strands often have sharp reorientations of the backbone and are accompanied by high curvature whereas the negative entries are usually those that enter from left-handed loop regions, as there is no need for the backbone to reverse the torsion. Excluding several kinks within β-strands, the exits lead either smooth into the next loop regions (in case of negative exits) or into well-defined turns. In the latter case, the exit is either positive or negative depending on the nature of the turn.

The discussion of the APSA results listed in Table 3 reveals that the number of α-helices and β-strands assigned by APSA is comparable to those suggested by existing methods such as DSSP, the disparities being further analyzed and found meaningful. APSA can also be used to quantify and systematically classify the regular as well as irregular structures leading to a more manageable and uniform structure description system, as all conformations are analyzed in the same way when they are classified. Turns are more variable among the secondary structures and owing to their non-repeating regularity, they are difficult to describe and categorize. It was shown in an earlier study [14] that turns that are similar (different) in 3D, indeed have similar (different) κ(s)-τ(s) patterns. The detail present in the κ(s)-τ(s) plots can be used to analyze kinks and distortions, which is sufficient proof that they contain extensive information regarding the direction and structure of turns. Thus, an analysis of a single protein is undertaken in Section 3.4 to show that the span of α-helices and β-strands as well as the nature of all loops and turns is accurately described by APSA.

### 3.4 APSA Description of Ubiquitin

The results of the application of APSA to ubiquitin (1UBQ) are summarized in Table 4. Ubiquitin [30] is an *alpha-and-beta* class protein with a *roll* topology according to the CATH [7] classification. It is a single chain protein with 76 amino acids that assume approximately 14 recognizable secondary structures, including an $\alpha$-helix, two short helical segments, five $\beta$-strands, and six turns. Table 4 compares the APSA assignment of the structural units of 1UBQ (for $\kappa(s)$ and $\tau(s)$ plots see Figure 2a) with i) a H-bonding- and $\phi$, $\psi$-based method used by Vijayakumar, Bugg, and Cook (VBC) [30], ii) the H-bonding-based DSSP method, [10] and iii) the secondary structure assignment used for the description of 1UBQ folding. [31] The terminology used in Table 4, assignment criteria, and the number of amino acids (span) of each structure are as stated in the original literature. [10,30-32] For example, the type III turns, as assigned by VBC, [30] have been well defined in literature as turns that have repeating $\varphi,\psi$ values of -60°, -30°, identical with those of the $3_{10}$-helix. The type III' turn would be its mirror image. A '$\beta$-turn' (turn 1) refers to the turn connecting two successive antiparallel $\beta$-strands.

Table 4, Figures 2a and 2b

The terms used in connection with APSA are (if not discussed in the previous sections): i) '$\beta$-Trough (peak)', which is a single strongly negative (positive) $\tau$-trough (peak) of a $\beta$-strand; ii) 'helical segment', which is used when the segment is helical, but the exact conformation is not typically an $\alpha$-, $3_{10}$- or $\pi$-segment; iii) '$\beta$-conformation', which refers to the $\beta$-peaks (or extended peaks) occurring at the respective amino acids.

The $\alpha$-helix from I23 to E34 was identified unambiguously by all assignment methods, and so were the five $\beta$-strands. Of the two helical segments, the 38 to 40 (148 Å to 158 Å, Figure 2a) one was variously described as a turn, a $3_{10}$-helix, or a short helix whereas the $\kappa(s)$-$\tau(s)$ diagrams clearly indicate $3_{10}$ character. The second helical segment from 56 to 59 (right after turn 4 at the N-terminus). was described as type III turn by VBC; DSSP assigned a $\beta$-bridge, a turn, and a $3_{10}$-helix in succession whereas the folding analysis considered two turns followed by a 'short helix'. As can be seen from the $\kappa(s)$ and $\tau(s)$ diagrams (Figure 2a), the region can be split in any of the ways mentioned. However, an accurate APSA-based description of this region is that amino acids 52 and 53 of turn 4 form a loop and then extend into 54 and 55 where a $3_{10}$-helix starts from the latter amino acid.

It is noteworthy that APSA is able to recognize single ß-peaks (troughs) for DSSP's 'isolated $\beta$-bridges', although this is not the topic of this investigation because loop regions are not analyzed here. These are examples of the effects of tertiary structure on secondary structure.

The β-bridge H-bond imposes the 'β-peak' conformation on the isolated amino acid as reflected in the τ(s) diagram (at 215 Å, Figure 2a). Among other proteins of the dataset though, this peak was found alongside other neighboring β-peaks leading to continuous ß-strand assignments. The fact that turns can be viewed as combinations of extended and helix conformations has been documented [1]. This feature is seen in several of the turn segments. In the 1UBQ segment from $s$ = 68 to 84 Å (Figure 2a), the P19 peak in τ is helical (compare with the first peak of the helix at 84 Å) whereas the other three amino acids have β-peaks. Similar features are recognizable with the other turns. The entry (and exit) of the polypeptide chain into (and out of) the helix at threonin 22, proline 37, (asparagine 60), etc. due to local unwinding results in the characteristic β-peaks. In addition, the way it reorients its general direction using glycines 10, 35 and asparagine 60 $C_{\alpha}$ atoms as pivots have been shown (Table 4, Figure 2a). The advantage of a graphical representation is exploited to visualize that the β-strands of 1UBQ, as seen from its κ(s) and τ(s) patterns, are not perfectly flat (compare with ideal β-strand in Figure 1n).

It can be seen that there are differences in structure assignment among the various methods. These differences arise not only due to the difference in the criteria used for assignment, but also due to the differing sensitivities in detecting the boundaries of the secondary structures. Early, it has been documented [33] that "ambiguity" is an intrinsic property of the protein, especially with respect to the turn regions that connect the boundaries of adjacent secondary structures (see Table 4). However, the similarity of turn 2 to turn 4 and its difference from β-turn-1 gives an idea to construct turn templates for loop regions. With respect to the choice of criteria, it should be remembered that the definition of the H-bond according to DSSP with respect to energy and distance is arbitrary and that the φ-ψ angle description of the polypeptide chain backbone is both discrete and local. As stated above, the deviation of the β-strands from the ideal is explicit and recognizable, especially as it is represented graphically. One can also relate to the specific parts of the secondary structure that is likely to deviate from the ideal. For example, the lysine 33 in the α-helix deviating from the rest of the helix that stretches from isoleucine 23 to glutamate 34 is evident from the κ(s) and τ(s) diagrams (compare ideal structure in Figure 1a).

Though the κ and τ information in Table 3 is mainly about α-helices and β-strands, it is well known that many more intermediate structures exist to allow many conformations to occur (among the loop regions). Analysis and classifications of these regions will be the topic of a forthcoming paper [35]. With APSA, longer loop regions can be quantitatively described as having helical and extended regions alone.

**3.5 Recognizing Common Architectures – An APSA Similarity Test**

Similar structural features have similar patterns in the κ(s),τ(s) diagrams. Figure 2b shows the τ(s) diagram of spinach ferredoxin 1A70, an iron-sulfur protein. It is 97 amino acids long and has the same *roll architecture* as ubiquitin (1UBQ, 76 residues) by CATH [7] classification. In the κ(s) and τ(s) diagrams of Figure 2b, the 2 helices, 5 β-strands, and 6 turns that resemble 1UBQ are indicated to aid comparison (see also Figure 3). Since torsion τ is an important and highly sensitive parameter, it is sufficient to use just τ for the comparison of 1A70 and 1UBQ.

Inspection of the τ (s) diagrams in Figure 2 immediately reveals the similarity of the two protein structures with regard to β-strands 1, 2, 3, 5, and helix 1. This can also be concluded when comparing the ribbon diagrams in Figure 3. However, the APSA diagrams of Figure 2 also reveal (dis)similarities in the non-regular structures such as the turns. For example turn 1 in 1UBQ is much more (right-left) twisted (larger ±τ-values, Figure 2a) than that in 1A70 (Figure 2b). The same applies to turn 2. Protein 1A70 has 2 additional features labeled 'helical turn segments 4 and 7', which differ from turns 4 and 7 in 1UBQ (Figures 2a and 2b). These loop regions are only slightly helical and account for the fact that 1A70 is longer. The *helical segment* of 1UBQ at $s$ = 220 Å (which is barely one turn of a $3_{10}$-helix; see curvature diagram in Figure 2a) is longer and α-helical in 1A70 (labeled *α-helix 2*), occupying approximately an equivalent 3D position. The short and crooked *β-strand 4* is found in both proteins, but is arranged differently in the sequence of secondary structural elements with respect to *α-helix 2*.

Table 5, Figure 3

In Table 5, regular and non-regular structures of the two proteins are compared by complementing the APSA information from Figure 2 (and Table 4) by appropriate 3D pictures. The similarities of turns 1, 2, 6, and the ß-conformation as reflected by the τ(s) diagrams are confirmed by the 3D-pictures (see comments in Table 5). In summary the τ(s) diagrams (optionally complemented by the κ(s) diagrams) provide a rapid, accurate, and detailed analysis of the structures of the two proteins, which is confirmed by appropriate ribbon diagrams.

**3.6 Comparison of Domain Similarity**

A fully automated and accurate method that can compare and classify proteins and protein segments at secondary, supersecondary, and tertiary levels without the need for manual intervention is not yet available. Though there are several databases of classified structures based on the proteins deposited in the PDB [6] such as CATH [7], SCOP [8], Dali [34], TOPS [36], etc., each of them uses a different approach to judge similarity among proteins. CATH and SCOP databases need manual analysis to complete the judgment of similarity. The update is sometimes accompanied by a rearrangement of previously classified structures when new structures are included into the database. The TOPS database makes the overall connectivity and folds visible by a rather drastic simplification of representing the secondary structures as cartoons. These methods are rigid in their assignment of secondary structures in the way that once a structure does not satisfy any of the limited definitions, the entire region is treated as 'loop.' Without further attempt to characterize the geometry in these regions, they are simply compared with the aim of getting differences.

In the light of the above need of having a more efficient and meaningful protein structure comparison method, it can be shown that in order to compare domains, averaging and simplification could be done *without the loss of details at the secondary level*. A direct comparison of the κ(s), τ(s) diagrams of 2 proteins of the same architecture (Section 3.5) reveals how domain similarity can be ascertained by the locations of the secondary structures and the overall similarity of the turns. The closer the folds of the two proteins, the more identical their κ-τ patterns become.

For the purpose of providing further proof for the fact that APSA is perfectly suited to quantitatively determine the (dis)similarity of protein structure, different domains are compared in the following way. A set of 20 domains was selected from the "*all alpha*" class, 15 belonging to the "*orthogonal bundle*" and 5 to the "*up and down Bundle*" and these were compared with each other. The CATH tree is shown and numbered in Figure 4 representing a sampling at all levels of the CATH classification. As a measure of the relationship of these domains, an "order of relationship" was set up by counting (from right to left in Figure 4) the number of CATH nodes separating two domains. Two domains of 0-order relationship belong to the same 'I level'; the 'D' level, the final level of CATH containing identical proteins, is not considered in the order scale. The highest number in terms of order is 8, signifying different classes. Thus, fourth order relationship domains belong to the same homology, but not to the same 'S level.' This class contains identical proteins and therefore is not considered in the order scale.

Table 6, Figures 4 and 5

As a measure of similarity, a grading scheme was set up with letters ranging from A (identical) to F (dissimilar) signifying decreasing similarity of domains within the same *all alpha* class (see Figures 4 and 5). The criteria A to F used were based on the number and ordering of secondary structures, κ(s), τ(s) patterns of the turns, types of entries and exits, the nature of loop regions, the size of the domains, and the overall ordering of the secondary structures with respect to each other (see Table 6). Allowance was given for variation; for example, some loop regions that appeared to be distorted helices were recognized similar to an α-helix (Table 6). A correlation of this *similarity* index was combined with the *order* index creating a similarity matrix (Figure 5). It is to be expected from such a correlation, that the smaller the order, the closer the relationship of the domains by CATH, the higher should be the grade of similarity assigned.

For an 'A' grade similarity of two domains (Figure 5) 99% of all amino acids have to have similar τ(s) patterns according to the properties listed in Table 6. An example is shown in Figure 6a where the τ(s) values of domains 1 and 2 having an order of relationship of 2 are identical. A reference to the length of the domain is made to accommodate greater flexibility in the longer loops of larger domains (Table 6), as in the case of domain 16, 17 and 18 that are about 300 amino acids long. Distortions in helices are permitted along with some minor variations. A grade 'B' similarity (Table 6, Figure 5) implies stronger distortions in secondary structures and/or differences in parts of turns such as 2 - 3 negative τ-troughs instead of positive ones. Domain 6 differs from domain 2 at amino acids 18-20, at the C-terminus of the last helix and at the arrangement of the last few amino acids. Stronger distortions that evidently bend helices to orient them differently in 3D space are graded with a 'C' similarity (Table 6), which also includes significant differences in the turns and loops owing to the different sizes of the domains being compared. In the case of domains 2 and 8, the first 3 helices of domain 2 strongly resemble the whole of domain 8. Thus, even though both domains are "3 helix bundles", the presence of extra helices in domain 2 can be clearly seen.

Figure 6

A grade that would interpret as "different" is 'D' (Table 6, Figure 5). When given a grade 'E', the secondary structures of the domains are present in totally different supersecondary arrangement making the fold of the domain significantly different. However, similarities can be seen between different parts of the protein. Some of the supersecondary structures are similar; however, they occur in a "jumbled" order, thus differing in topology. A greater difference leads to a grade of 'F'. It is noteworthy that, though the domain as a whole (boundaries as prescribed by CATH) is considered to be *very different* by this index, similar supersecondary structures and folds can still be recognized at various parts. For example, among the first few helices of domain 17 (1socA02) ranging approximately from amino acid 536 to 644, several secondary structure and turn features can be identified belonging to the *orthogonal bundle* architecture. A comparison with domain 9 as shown in Figure 6b clarifies the fact that though the loop in domain 17 is more meandering resulting in the many oscillations in $\tau(s)$, the patterns are equivalent. The two positive β-peaks at 221 and 222 in 1YOV correspond to the same at 549 and 551 in 1S0C. As discussed in Section 3.3, one β-peak (at 244, 1YOV) is equivalent, by rotation to four β-peaks (at 552, 1S0C). After accounting for these rotations and translations, the equivalence of the 2 segments can be seen in the 3D inset. The long helices and the turns that appear after amino acid 644 in 1S0C, however, clearly reflect a different arrangement, namely the *up-and-down bundle*. As the similarity assignment from A to F is done only for domains within the same class (*all alpha*), greater differences that occur beyond the *all alpha* class of domains are not documented.

## 4. Conclusions

The performance of APSA being based on the determination of curvature and torsion of the protein backbone has been demonstrated in this work. Previous protein structure descriptions, which have taken an approach related in some way to APSA, have been discussed in Ref. 14 and the advantages of APSA with regard to these approaches have been worked out there and do not need to be repeated here.

A systematic analysis performed on 5 α helices and 8 β strands (Table 1) resulted in the derivation of a working definition for the same secondary structures in terms of curvature and torsion patterns, κ(s) and τ(s). An automated analysis of 77 proteins carried out with APSA led to a secondary structure assignment that was compared to that of DSSP. A total of 533 α-helices and 644 β strands were recognized by APSA, whereas DSSP's assignments (536 α-helices and 651 ß-strands) differed for 20 α-helices (12 more, 8 less) and 46 ß-strands (24 more, 22 less). Though the approaches are vastly different, the total number of structures was thus found comparable. In addition, the conformational features in 3D space were accurately described in the 2D κ(s) and τ(s) diagrams. From τ(s) alone, in most cases, kinks and distortions could be recognized and quantified. A list of distortions was also discussed as occurring in the body and termini of α-helices and β-strands. A way of describing distorted helical termini based on whether the diameter of the region was larger or smaller than the α-helix, as deduced from low or high κ(s) values and variations in τ(s) was presented.

Similar structural features between any two proteins also become evident in APSA's κ(s) and τ(s) diagrams. The *roll* architecture of ferridoxin (1A70) and ubiquitin (1UBQ) were compared. Two extra loop regions of the former protein between residues 32 to 44 and 57 to 64 that correspond to an increase in overall length of the fold were shown. In the wake of such a comparison, the degree of CATH relationship and index of similarity was correlated in an analysis that compared twenty *all alpha* domains with each other. It was shown that these 2D κ(s), τ(s) patterns could be used for similarity comparisons at any level whether secondary, super-secondary, or tertiary. Accordingly, domains of different homologous superfamily, topology, and architecture were shown to have increasingly different κ and τ profiles.

The APSA method accurately reflects the conformation of the backbone effectively reducing 3D information to a 2D representation. The method is mathematically well founded and computationally robust, describing each secondary structure with a unique κ(s), τ(s) pattern reflecting its 3D properties. Analysis of the 78 protein chains investigated in this work with APSA requires about 1 sec computer time. Hence, APSA is well-suited for the rapid structure analysis of the 50,000 proteins of the PDB. It provides a complete conformational analysis and identification of all residues of a protein.

It is a continuous representation where a global trend in conformation can be seen for all amino acids, whether they are in the helical, extended or loop regions of proteins. The speed and the simplicity of the analysis is due to the use of a simplified backbone representation. It was demonstrated that APSA can be easily applied to the analysis of supersecondary and tertiary structure. [45]

APSA is exclusively based on conformational (structural) protein data as reflected by the positions of the $C_\alpha$ atoms in the protein backbone whereas the DSSP description strongly depends on the types and arrangements of H-bonding in the protein. APSA does not need any charge or energy information, which are essential for DSSP. This is a clear advantage over DSSP's assessment of backbone structure since H-bonding patterns do not supply information on the distortions and orientations of backbone structures. Otherwise, APSA and DSSP should complement each other where APSA should take the lead in the structural analysis because of its rapid description and DSSP should come in with additional information, especially on H-bonding.

**Supporting Information**: A table with the 77 proteins, their names, and PDB identification numbers is given in the Supporting Information. Also, curvature and torsion diagrams, $\kappa(s)$ and $\tau(s)$, and the backbone line are listed for each protein investigated.

## Acknowledgements

DC and EK thank the University of the Pacific for support. Support by the NSF under grant CHE 071893 is acknowledged

## References

1. Richardson JS. The anatomy and taxonomy of protein structure. Adv in Prot Chem. 1981; 34:167-339.
2. Chelvanayagam G, Heringa J, Argos P. Anatomy and evolution of proteins displaying the viral capsid jellyroll topology. J Mol Biol 1992;228:220-242.

3. Scheerlinck JPY, Lasters I, Claessem M, DeMaeyer M, Pio F, Delhause P, Wodak SJ. Recurrent αß loop structures in TIM barrel motifs show a distinct pattern of conserved structural features. Proteins: structure, function, and genetics 1992;12:299-313.

4. Murzin AG, Finkelstein, AV. General archtecture of the alpha-helical globule. J Mol Biol 1988;204:749-769.

5. Richardson, JS. ß-Sheet topology and the relatedness of proteins. Nature 1977;268:495-500.

6. Berman HM, Westbrook J, Feng Z, Gilliland G, Bhat TN, Weissig H, Shyndalyov IN, Bourne PE. The Protein Data Bank 2000, Nuclei Acid Res. 2000;28:235-242.

7. Orengo CA, Michie AD, Jones S, Jones DT, Swindells MB, Thornton JM. CATH- A Hierarchic Classification of Protein Domain Structures. 8. Structure 1997;5:1093-1108.

8. Murzin AG, Brenner SE, Hubbard T, Chothia C. SCOP: a structural classification of proteins database for the investigation of sequences and structures. J Mol Biol 1995;247:536-540.

9. Frishman D, Argos P.Knowledge-based protein secondary structure assignment. Proteins 1985;23:566-579.

10. Sanger K. Dictionary of secondary structure of proteins: pattern recognition of hydrogen-bonded and geometrical features. Biopolymers 1983;22:2577-2637.

11. Richards FM, Kundrot CE. Identification of structural motifs from protien coordinate data: secondary and first level supersecondary structure. Proteins 1988;3:71-84.

12. Levitt M, Greer J. Automatic identification of secondary structure in globular proteins. J Mol Biol 1977;114:181-293.

13. Venkatachalam C, Stereochemical criteria for polypeptides and proteins: conformation of a system of three linked peptide units Biopolymers 1968;6:1425-1436.

14. Ranganathan S. Izotov D, Kraka E, Cremer D, Automated and accurate protein structure description: Distribution of Ideal Secondary Structural Units in Natural Proteins, arXiv: 0811.3252v2 [q-bio.QM].

15. Arab S, Didehvar F, Eslahchi C, Sadeghi M. Helix segment assignment in proteins using fuzzy logic. Iranian J of Biotech. 2007;5:93-99.

16 Humphrey W, Dalke A, Schulten K. VMD - Visual Molecular Dynamics, J. Molec. Graphics 1996;14:33-38.

17. Benedetti E, Blasio BD, Pavone V, Pedone C, Santini A, Crisma M, Toniolo C. Molecular Conformations and Biological Interactions: The 3-10 and alpha-helical conformation in peptides. Indian Academy of Sciences, 1991, p. 497-502.
18. Barlow DJ, Thornton J M. 3Helix geometry in proteins. J Mol Biol 1988;201:601-619.
19. Kumar S, Bansal M, Velavan R. HELANAL: a program to characterize helix geometry in proteins. J Biomol Struct Dyn 2000;17:811-819.
20. Cartailler JP, Luecke H. Structural and functional characterization of pi bulges and other short interhelical deformations. Struct (Camb) 2004;12:133-144.
21. Sun Z, Blundell T. The patter of common supersecondary structure (motifs) in protein databases. Proceedings of the 28th Annual Hawaii International Conference on System Sciences, 1995:312.
22. Sun Z-R, Zhang C-T, Wu F-H, Peng L-W. A vector projection method for predicting supersecondary motifs. J Protein Chem 1996;15:721.
23. Taylor H. Proc Am Phyl Soc 1941;85:1.
24. Doig AJ, MacArthur MW, Stapely BJ, Thormton JM. Structure of N-termini of helices in proteins. Protein Science 1997;6:147-155.
25. Karpen ME, De-Aseth PL, Neet KE. Differences in amino acid distribution of 3[10]-helices and alpha-helices. Protein Science 1992;1:1333-1342.
26. Toniolo C, Crisma M, Formaggio F, Peggion C, Broxterman Q, Kaptein B. Peptide β-bend and $3_{10}$ helix: from 3D structural studies to applications as templates. J Inclusion phenomena nad macrocyclic chemistry 2005;51:121-136.
27. Fodje MN, Al-Karadaghi S. Occurance, conformational features and amino acid propensities for the pi-helix. Protein Eng 2002;15:533-358.
28. Nelson, DL, Cox, MM. Lehninger Principles of Biochemistry, 5th Edition, Freeman, New York, 2008.
29. a) Sowdhamini R, Srinivasan N, Ramakrishana C, Balaram P. Orthogonal ββ motifs in proteins. J. Mol. Biology 1992;223:845-851. b) Efimov, AV. Standard structures in proteins. Prog Biophys Molec Biol 1993;60:201-239.
30. Vijayakumar S, Bugg C E, Cook W J. Structure of ubiquitin refined at 1.8 A resolution. J Mol Biol 1987;194: 531-544.

31. Multiple folding mechanisms of protien ubiquitin. Jian Zhang, Meng Quin, Wei Wang. 3, 2005, Proteins: structure, function and bioinformatics, Vol. 59, pp. 565-579.

32. Protein Data Bank. www.rcsb.org/pdb. [Online]

33. Rose GD, Gierasch LM, Smith JA. Turns in peptides and proteins. Adv Prot Chem 1985;37:1-109.

34. Holm L, Sander C. Mapping the protein universe. Science 1996;273:595-603.

35. Rackovsky S, Scheraga HA. Differential Geometry and Polymer Conformation .1. Comparison of Protein Conformations. Macromolecules 1978;11: 1168-1174.

36. Louie AH, Somorjai RL. Differential Geometry of Proteins Helical Approximations. J. Mol. Biol. 1983;168:13-162.

37. Louie AH, Somorjai RL. Differential Geometry of Proteins - A Structural and Dynamical Representation Of Patterns. J. Theor. Biol. 1982;98:189-209.

38 Soumpasis DM, Strahm MC. Efficient identification and analysis of substructures in proteins using the kappa-tau framework: Left turns and helix c-cap motifs. J. Bimolecular Structure and Dynamics 2000;17:965-979.

39. Hausrath AC, Goriely A. Continuous representations of proteins: Construction of coordinate models from curvature profiles. J. Struct. Biol. 2007;158:267-281.

40. Sklenar H, Etchebest C, Laverey R. Describing Protein-Structure - A General Algorithm Yielding Complete Helicoidal Parameters And A Unique Overall Axis. Proteins: Structure, Functions, and Genetics 1989;6:46-60.

41. Zhi D, Krishna S, Cao H, Pevzner P, Godznik A. Representing and comparing protein structures as paths in three-dimensional space. BMC Bioinformatics, 2006;7:460-472.

42. Can T, Wang YF. Protein Structure Alignment And Fast Similarity Search Using Local Shape Signatures. J. Bioinformatics and Comp. Biol. 2004;2:215-239.

43. Goyal S, Perkins NC, Lee CL. Nonlinear dynamics and loop formation in Kirchhoff rods with implications to the mechanics of DNA and cables. J. Comp. Phys. 2005;209:371-389.

44. Goyal S, Lillian T, Blumberg S, Meiners JC, Meyhofer E, Perkins NC. Intrinsic curvature of DNA influences LacR-mediated looping. Biophys. J. 2007;93: 4342-4359.

45. Ranganathan S. Izotov D, Kraka E, Cremer D, Projecting Three-dimensional Protein Structure into a One-dimensional Character Code Utilizing the Automated Protein Structure Analysis Method. arXiv:0811.3258v2 [q-bio.QM].

## Figure Captions

**Figure 1**. Curvature (above) and torsion diagrams (below), κ(s) and τ(s), for typical secondary structural units given in form of ribbon presentations. Every peak in a diagram reflects the conformation of a residue in the analyzed segments. (**a**) Ideal 14-residue long polyalanine α-helix. (**b**) Natural α-helix with small irregularities (1U4G: residues L131 to Y155). (**c**) A kink in an α-helix (1V54: P77-G97). (**d**) Distortions of an α-helix leading to a looser N-terminus (1QOY: E18-P36). (**e**) A strong kink leading to a large τ-value in a slightly distorted α-helix (2CPP: S258-G276). (**f**) Difference between body and N-terminus of an α-helix (1TVF: N72-S95). (**g**) An $\alpha_\pi$ cap at the C-terminus of an α-helix leading to higher curvature (5MBN: G5-D20). (**h**) Transition from a helix to a turn region with gradually increasing curvatures and interspersed high torsions (1QTE: W17-L37). (**i**) C-terminal caps of α-helix (1RWZ: I5-I27). (**j**) N-terminal caps of α-helix (1KSS: N543-F556). (**k**) N-terminal caps of $3_{10}$-helix (1QAZ: G74-L97). (**l**) C-terminal caps of $3_{10}$-helix (1MG6: Q93-S116). (**m**) N-terminal caps of α-helix with positive entry (1CTQ: G12-N26). (**n**) Ideal 4-residue long polyalanine ß-strand. (**o**) Distorted ß-strand (1RIE: H161-L178). (**p**) Degree of rotation of a second helix with regard to a first as reflected by the number of β-troughs in the τ-diagram of the connecting turn (1CDP 1-19 to 1HMD 55-66 to 1CTF 67-90 (below)); each additional β-trough indicates a left-handed 90°-rotation of the second helix.

**Figure 2.** Torsion and curvature diagrams, κ(s) and τ(s), of (**a**) ubiquitin (1UBQ) and (**b**) spinach ferridoxin (1A70) also having the *roll* architecture of ubiquitin. Structural regions obtained from APSA are separated by vertical dashed lines and identified by a short term. Compare with the ribbon diagrams of 1UBQ and 1A70 given in Figure 3. See text and Table 4 (Table 5) for more details.

**Figure 3.** Ribbon diagrams of 1UBQ (top) and 1A70 (bottom).

**Figure 4.** CATH [7] similarity relationships for 20 domains that are shown on the far right. The CATH levels are given on the bottom and the orders of relationship on the top. See text for details.

**Figure 5.** A similarity matrix constructed for 20 domains (see Figure 4) with the order of relationship taken from CATHSOLID classification system on the upper half of the matrix and the graded APSA similarity on the lower half. The order of relationship is given by the number of nodes separating the domains as counted from the relationship chart shown in Figure 4. Note that E* is between D and E.

**Figure 6. a)** Domains 1 (1a6i001) and 2 (2tct001) both ranging from amino acids 2 to 66 of the respective proteins whose $\tau$ are identical; *order of relationship* = 2, *similarity index* = A. **b)** A comparison of the orthogonal helix pairs from domains 9 (1yovB02) and 17 (1s0cA02) show resemblances in $\tau$ and 3D arrangement. See text for details.

**Scheme 1**. The four windows W1, W2, W3, and W4 are schematically shown presenting the curvature $\kappa(s)$ and torsion ranges of $\tau(s)$ for each Wn, the peak (trough) forms, and the terms used in the text for describing the windows.

**Scheme 2**. A hypothetical helix with an $\alpha$-helix body and two caps at either end is shown. The helix is started by the ‘starter’, ended by the ‘exit’ and the ‘entry’ is defined by the residue just prior to the starter. A distinct 3D structure formed by the few residues into the helix from either end is termed as a “terminus” by APSA. On the left side the terms are given that are given in the literature [1] where helix entry and helix exit are added according to Efimov. [29a] The APSA terminology tries to follow the terminology used in the literature however considers at the same time the exact definition of terms via curvature and torsion.

**Table 1.** Determination of working ranges for curvature κ(s) and torsion τ(s) using 5 α-helices and 8 β-strands with 10 equidistant spline points between every $C_\alpha$ atom. [a]

| **A. α-Helices** | | | | | | | |
|---|---|---|---|---|---|---|---|
| PDB ID | Helix position | κ | | | τ | | |
| | | avg | min | max | avg | min | max |
| 1A6I | 128-148 | 0.39 | 0.29 | 0.59 | 0.15 | 0.08 | 0.22 |
| 1BJZ | 130-148 | 0.39 | 0.29 | 0.60 | 0.15 | 0.07 | 0.23 |
| 1R4M(B) | 23-29 | 0.40 | 0.30 | 0.62 | 0.15 | 0.08 | 0.21 |
| 1R4M(B) | 426-439 | 0.40 | 0.29 | 0.62 | 0.14 | 0.07 | 0.21 |
| 1S0D | 589-604 | 0.39 | 0.29 | 0.66 | 0.15 | 0.07 | 0.23 |
| Overall | 76 | **0.39** | **0.29** | **0.66** | **0.15** | **0.07** | **0.23** |

| **B. ß-Strands** | | | | | |
|---|---|---|---|---|---|
| PDB ID | Strand position | κ(min) | κ(max) | τ (min) | τ (max) |
| 1UYL | 78-81 | < 0.07 | 0.76 – 1.3 | < -2.1 | -0.03 to -0.08 |
| 1UYL | 89-93 | < 0.06 | 0.63 – 1.2 | < -1.8 | -0.01 to -0.07 |
| 1ITV | 18-21 | < 0.09 | 0.85-1.0 | < -1.8 | -0.09 to -0.07 |
| 1ITV | 26-31 | < 0.18 | 0.5-0.9 | < -1.2 | -0.11 to -0.004 |
| 1ITV | 74-78 | < 0.06 | 0.5-1.0 | < -2.8 | -0.08 to -0.006 |
| 2PCY | 25-30 | < 0.14 | 0.7-1.3 | < -1.2 | -0.12 to -0.07 |
| 2PCY | 37-42 | < 0.18 | 0.5-1.2 | < -0.8 | -0.15 to -0.05 |
| Overall | | **< 0.14** | **0.4-1.3** | **< -0.8** | **-0.15 to -0.004** |
| 1V86 | 2-6 | **< 0.11** | **0.48-0.7** | **> +0.97** | **+0.02 to +0.14** |

[a] Protein structures investigated are given by their PDB identification (ID) number and the residue numbers. The terms *min* and *max* denote the smallest and largest κ(s) or τ(s) values, *avg* the average of all 10 values calculated.

**Table 2.** Specification of $\kappa(s)$, $\tau(s)$ windows for helices and ß-strands used for automation and comparison with ideal helices and extended structures. [a]

| **Automation details**<br>**Windows Wn** | **Working values**<br>**[$Å^{-1}$]** | **Ideal values**<br>**[$Å^{-1}$]** | **Minimum length**<br>**[# of residues]** |
|---|---|---|---|
| W1: $\alpha$-Helix (starter residue) | $0.2 \leq \tau(\max) \leq 0.4$ | $0.3 \leq \kappa \leq 0.56$<br>$0.08 \leq \tau \leq 0.18$ | 4 |
| W2: $\alpha$-Helix; body | $0.25 \leq \kappa \leq 0.67$<br>$0.05 \leq \tau \leq 0.24$ | | |
| W3: $\beta$-Strand: negative $\tau$: troughs<br>(left-handed) | $0.5 \leq \kappa \leq 1.4$<br>$0.0 \leq \kappa(\min) \leq 0.02$<br>$-0.001 \leq \tau(\max) \leq -0.15$<br>$\tau(\min) < -0.75$ | $0.01 \leq \kappa \leq 1.0$,<br>$\tau(\min) < -2.9$ | 3 |
| W4: $\beta$-Strand: positive $\tau$: peaks<br>(right-handed) | $0.4 \leq \kappa(\max) \leq 0.9$<br>$0.0 \leq \kappa(\min) \leq 0.02$<br>$0.001 \leq \tau(\min) \leq 0.15$<br>$\tau(\max) > 0.75$ | $0.01 \leq \kappa \leq 1.0$,<br>$\tau(\max) > 2.9$ | 3 |

[a] The terms *min* and *max* denote the smallest and largest $\kappa(s)$ or $\tau(s)$ values in the range from one $C_\alpha$ atom to the next higher $C_\alpha$ atom.

**Table 3.** APSA results for a dataset of 77 proteins. [a]

| CATH Architecture | S. No | PDB ID | # of α DSSP | # of α APSA | Comments | # of β DSSP | # of β APSA | Comments |
|---|---|---|---|---|---|---|---|---|
| **CLASS: Mainly alpha** | | | | | | | | |
| α Horse-shoe | 1 | 1M8Z(A) | 28 | 28 | | 0 | 0 | |
| | 2 | 1QTE(A01) | 37 | 37 | 1(split) | 0 | 2 | 2strand+ |
| | 3 | 1V54(E) | 5 | 5 | 1(split) | 0 | 1 | 1strand+ |
| α Solenoid | 4 | 1PPR(M) | 16 | 16 | 2(split) | 0 | 0 | |
| α/α Barrel | 5 | 1QAZ(A) | 12 | 12 | | 0 | 2 | 2strand+ |
| Orthogonal bundle | 6 | 1ECA(A) | 8 | 8 | | 0 | 0 | |
| | 7 | 1GM8(B03) | 14 | 14 | 1(split) | 42 | 41 | 1strand- |
| | 8 | 1HC0(A) | 7 | 7 | | 3 | 3 | |
| | 9 | 1KSS(A02) | 20 | 20 | 1(split) | 19 | 19 | |
| | 10 | 1LMB(3) | 5 | 5 | 1(split) | 0 | 0 | |
| | 11 | 1NG6(A) | 7 | 7 | | 0 | 0 | |
| | . | 1QTE(A02,3) | | | | | | |
| | 12 | 1U4G(A02) | 8 | 8 | | 10 | 10 | |
| | 13 | 1UTG(A) | 4 | 4 | | 0 | 0 | |
| | 14 | 2CPP(A) | 19 | 18 | 2(split), 1(distort)- | 14 | 12 | 2strand- |
| | 15 | 2CTS(A) | 20 | 20 | 4(split) | 2 | 2 | |
| | 16 | 2LZM(A) | 10 | 10 | 1(split) | 3 | 3 | 1strand+, 1strand- |
| | 17 | 2MHB(A) | 8 | 8 | 1(split) | 0 | 0 | |
| | 18 | 3WRP(A) | 6 | 6 | | 0 | 0 | |
| | 19 | 5MBN(A) | 9 | 9 | | 0 | 0 | |
| | 20 | 5PAL(A) | 7 | 7 | | 2 | 2 | |
| Up and down bundle | 21 | 17GS(A02) | 10 | 10 | | 4 | 4 | |
| | 22 | 1AA7(A) | 10 | 9 | 1(distort)- | 0 | 0 | |
| | 23 | 1MG6(A) | 4 | 4 | | 2 | 2 | |
| | 24 | 1O83(A) | 6 | 6 | | 0 | 0 | |
| | 25 | 1QOY(A) | 9 | 8 | 1(distort)- | 2 | 2 | |
| | . | 1V54(A) | 21 | 20 | 3(split), [2] | 1 | 1 | |
| | 26 | 1VKE(B) | 5 | 5 | | 0 | 0 | |
| **CLASS: Mainly beta** | | | | | | | | |
| 3 Solenoid | 27 | 1EZG(A) | 0 | 0 | | 6 | 6 | |
| | 28 | 1QRE(A) | 2 | 2 | | 22 | 24 | 2strand+ |
| 3 Layer Sandwich | 29 | 1NYK(A) | 1 | 1 | | 11 | 9 | 1strand-, [2strand] |
| | 30 | 1RIE(A) | 1 | 2 | 1+ (:3-10) | 10 | 9 | 1strand- |
| 4 layer Sandwich | . | 1GM8(B01) | | | | | | |
| 4 Propeller | 31 | 1ITV(A) | 3 | 4 | 1+ | 17 | 17 | |
| β Barrel | 32 | 1EY0(A) | 3 | 3 | | 8 | 7 | 1strand- |
| | 33 | 2POR(A) | 3 | 3 | | 16 | 16 | |
| | 34 | 4PEP(A) | 7 | 7 | | 24 | 21 | 3strand- |
| β Complex | 35 | 1AQ2(A02) | 16 | 17 | 1(:H-bo Turn)+ | 28 | 28 | |
| Ribbon | 36 | 1TGX(A) | 0 | 0 | | 5 | 5 | |
| Roll | 37 | 1G79(A) | 6 | 6 | | 10 | 9 | 1strand- |
| | 38 | 1GCQ(A) | 0 | 0 | | 5 | 5 | |
| | . | 1GM8(B02) | | | | | | |
| | 39 | 1TVF(A02) | 11 | 10 | 1(distort)- | 16 | 16 | |
| | 40 | 1ZX6(A) | 0 | 0 | | 5 | 5 | |
| Orthogonal prism | 41 | 1B2P(A) | 0 | 0 | | 12 | 12 | |
| Sandwich | 42 | 1GCS(A) | 1 | 1 | | 14 | 13 | 1strand- |
| | 43 | 1REI(A) | 0 | 0 | | 10 | 10 | |
| | 44 | 2AZA(A) | 2 | 2 | 1(cap)+, 1(distort)- | 8 | 8 | |
| | 45 | 2PAB(A) | 1 | 1 | | 9 | 9 | |
| | 46 | 2PCY(A) | 0 | 0 | | 8 | 8 | |
| | 47 | 2SOD(O) | 0 | 0 | | 9 | 8 | 1strand- |
| Single sheet | 49 | 7RXN(A) | 0 | 0 | | 3 | 2 | 1 strand- |
| Trefoil | 50 | 1WBA(A) | 0 | 0 | | 13 | 11 | 1strand- |
| | 51 | 2FGF(A) | 0 | 0 | | 10 | 10 | |
| **CLASS: Alpha and Beta** | | | | | | | | |
| 2 Layer Sandwich | 52 | 1B4V(A02) | 12 | 11 | 1(distort)- | 19 | 19 | 1(split)+ |
| | 53 | 1B8S(A02) | 12 | 11 | 1(distort)- | 19 | 19 | |
| | 54 | 1CF3(A03) | 17 | 16 | 1(distort)- | 20 | 20 | 2strand- |
| | 55 | 1CRN(A) | 3 | 3 | | 2 | 2 | |
| | 56 | 1CTF(A) | 3 | 3 | | 3 | 3 | |
| | 57 | 1TGSI(I) | NA | 1 | 1 helix + | NA | 2 | |
| | 58 | 2CI2(I) | 1 | 1 | | 3 | 4 | 1strand+ |
| 3 Layer (aba) Sandwich | . | 17GS(A01) | | | | | | |
| | 59 | 1CTQ(A) | 6 | 7 | 1(split), 1(cap)+ | 6 | 6 | |
| | . | 1TVF(A01) | | | | | | |
| | 60 | 1WPU(A) | 4 | 4 | | 4 | 4 | |
| | 61 | 2AK3(A) | 8 | 8 | 1(split) | 7 | 7 | |
| | 62 | 2FOX(A) | 5 | 6 | 1+ | 6 | 6 | |
| | 63 | 5CPA(A) | 9 | 9 | | 8 | 12 | 4strand+ |
| 3 Layer (bba) Sandwich | . | 1B4V(A01) | | | | | | |
| | . | 1B8S(A01) | | | | | | |
| | . | 1CF3(A01) | | | | | | |
| | . | 1KSS(A03) | | | | | | |
| | 64 | 3GRS(A) | 14 | 14 | | 23 | 23 | 2strand+, 2strand- |
| α-β Barrel | 65 | 1H61(A) | 11 | 11 | | 14 | 12 | 2strand- |
| α-β Complex | 66 | 1B8P(A02) | 12 | 11 | 1(distort)- | 14 | 13 | 1strand- |
| | 67 | 1F7L(A) | 4 | 4 | | 5 | 5 | |
| | . | 1KSS(A01) | | | | | | |
| | 68 | 2CDV(A) | 4 | 3 | 1(distort)- | 4 | 3 | 1strand- |
| | 69 | 3HSC(A) | 12 | 13 | 1(:H-bo Turn)+ | 18 | 19 | 1strand+ |
| | 70 | 7AAT(A) | 16 | 16 | | 13 | 13 | |
| | 71 | 9PAP(A) | 5 | 5 | | 8 | 8 | |
| α-β Horseshoe | 72 | 1OZN(A) | 2 | 1 | 1(distort)- | 18 | 17 | 1strand+, 1strand-, [2strand] |
| Box | 73 | 1RWZ(A) | 4 | 4 | | 18 | 19 | 1strand+ |
| Roll | . | 1U4G(A01) | | | | | | |
| | 74 | 1UBQ(A) | 1 | 2 | 1(:3-10) | 5 | 5 | |
| | 75 | 2CA2(A) | 4 | 3 | 1(distort)- | 15 | 15 | |
| | 76 | 2CAB(A) | 3 | 3 | | 16 | 16 | |
| **CLASS: Few secondary structures** | | | | | | | | |
| Irregular | . | 1CF3(A02) | | | | | | |
| | 77 | 1HIP(A) | 2 | 2 | | 3 | 4 | 1strand+ |
| | 78 | 5PTI(A) | 1 | 1 | | 2 | 4 | 2strand+ |
| | | | **547** | **543** | | **656** | **654** | |

[a] **PDB ID** denotes the Protein Data Bank Identifier including chain and domain IDs where appropriate. The symbols **# of α** and **# of ß** denote the number of α-helices and β-strands, respectively, recognized by either **DSSP** (Dictionary of Secondary Structure of Proteins [10]) or the APSA (curvature-torsion based) method. The secondary structures are also commented as being **split**, **distorted**, and labeled as a "hydrogen-bonded turn" **(H-bo Turn)** by DSSP; an α-helical cap **(cap)** is labeled as $3_{10}$ helix by DSSP **(:3-10)**. The **+ (-)** signs following each expression indicate that the secondary structure was added to (subtracted) from the total APSA count of α- or ß-structures. The symbol **[ ]** indicates that two secondary structures are merged by APSA.

**Table 4.** Comparison of the conformational (structural) features of ubiquitin (1UBQ) as described by APSA and three different other methods taken from the literature.

| Secondary Structure ranges | | | | Approx. **s** values APSA | | APSA Secondary structure | Comments [d] |
|---|---|---|---|---|---|---|---|
| VBC [a] (PDB)r | DSSP [b] | Folding analysis [c] | APSA (this work) | Start | End | | |
| M1-T7 | Q2-T7 | ~β-1 | Q2-T7 | 0 | 25 | β-Strand 1 | τ Plot: Regular patterns that resemble the ideal beta strand |
| T7-G10 (β-Turn) | L8-T9 | ~Turn 1 | L8-G10 | 25 | 38 | Turn 1 | τ Plot: Successive and even number of sign changes shows a flat turn region. |
| | | | G10 | 36 | | Glycine pivot | Sharp reorientation of backbone at G10 causes the steady sign change through 0 in τ and a strong κ. |
| G10-V17 | T12-E16 | ~β-2 | K11-E18 | 38 | 68 | β-Strand 2 | τ Plot: Regular patterns that resemble the ideal beta strand |
| E18-D21 | P19-S20 | ~Turn 2 | P19-T22 | 68 | 84 | Turn 2 | κ & τ Plot: Partial helix character<br>τ Plot: Sign changes. |
| | T22 (Turn) | | T22 | | | Helix entry | τ Plot: extended conformation [c] at helix entry |
| I23-E34 | I23-E34 | α-Helix | I23-E34 | 84 | 134 | α-Helix 1 | κ and τ Plot: End of helix shows slight distortion [e] |
| | | | G35 | 132 | | Glycine pivot | κ and τ Plot: Sharp reorientation as in G10. |
| | | | I36 | 134 | 148 | ß confor- | τ Plot: ß peak |
| | | | | | | mation | |
| | | | P37 | 144 | | Helix entry of the turn | τ Plot: extended conformation at helix entry as in T22. |
| P37-Q40 (Type III turn) | P38-Q40 ($3_{10}$ helix) | P38-Q40 (short helix1) | P38-Q40 | 148 | 160 | Turn 3 | κ & τ Plot: shows $3_{10}$ helix character of a Type III turn. |
| Q40-F45 | Q41-F45 | ~β-3 | Q41-I44 | 160 | 174 | β-Strand 3 | τ Plot: Short distorted segment. |
| F45-K48 (TypeIII' turn) | A46-G47 | ~Turn 3 | F45-G47 | 174 | 186 | Turn 4 | κ and τ Plot: turn has partial helix character. |
| K48-L50 | K48-Q49 | ~β-4 | K48-E51 | 186 | 202 | β-Strand 4 | κ Plot: Short and bent [e] |
| E51-R54 | D52-G53 | ~Turn 4, | D52-T55 | 202 | 220 | Turn 5 | 1. Mixed helix and β-character in κ & τ peaks, as in Turn 2.<br>2. Extended conformation (helix entry) at T55, as in T22 (τ-Plot)<br>3. T55-D58 region shows slight $3_{10}$ character of Type III turn (κ-Plot) |
| | T55 (β-Bridge) | | | | | | |
| T55-D58 (Type III turn) | | ~Turn 5 | | | | | |
| L56-Y59 | L56 (Turn) | L56-Y59 (Short helix2) | L56-Y59 | 220 | 234 | Helical segment | DSSP's $3_{10}$-helix is an α-helix by torsion and a $3_{10}$ helix by κ. |
| | S57-Y59 ($3_{10}$helix) | | | | | | |
| | N60 (Turn) | | N60 I61 | | | ß-confor-mation | Very strong κ at N60 shows strong bending of the backbone. |
| Q62-S65 | K63-E64 | ~Turn 6 | Q62-E64 | 243 | 258 | Turn 6 | κ Plot: Partial helix character.<br>τ Plot: Sign changes. |
| E64-R72 | T66-L71 | ~β-5 | S65-L73 | 258 | 283 | β-Strand 5 | κ Plot: Long and relatively flat. [e] |
| | | ~Turn 7 | R72 | 284 | end | Turn 7 | κ Plot: R72 shows strong κ within the β-strand. |

[a] Ref. 30. [b] Ref. 10. [c] Ref. 31. [d] The term β (or extended) conformation has been used based on this work; it refers to the conformation of the individual amino acid (as taken up in a β-strand) as determined from κ(s)-τ(s) plots. – Turns are given according to Ref. [5]. Turns can be viewed as combinations of helix- and strand-amino acid conformations [5] and this is reflected in the κ(s)-τ(s) plots. [e] Comparison of κ patterns to the ideal (Figure1n) shows slight strand distortion.

**Table 5**. Some regular and non-regular structural features of spinach ferredoxin (1A70) compared with those of ubiquitin (1UBQ). [a]

| 1UBQ features<br><br>APSA | Pictures | 1A70 features<br><br>APSA | Pictures |
|---|---|---|---|
| Turn 1<br><br>(s = 25 to 38) | 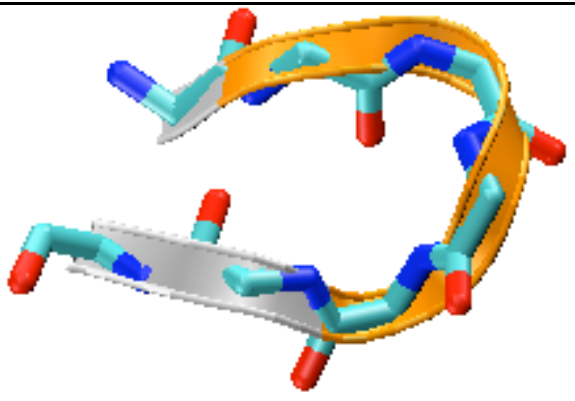 | **P10-G12** More planar than turn 1 in 1UBQ: the τ averages to 0.<br><br>(s = 33 to 46) | 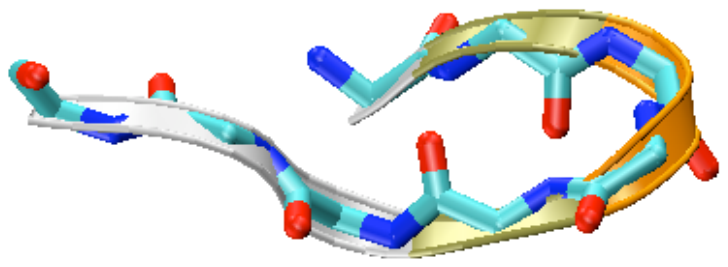 |
| Turn 2<br><br>(s = 68 to 84) | 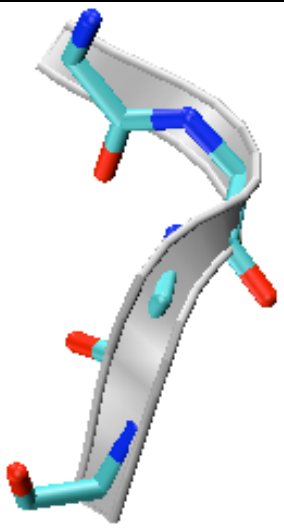 | **D20-Y23** The first part of the turn is partly helical and resembles the turn in 1UBQ<br><br>(s = 72 to 88) | 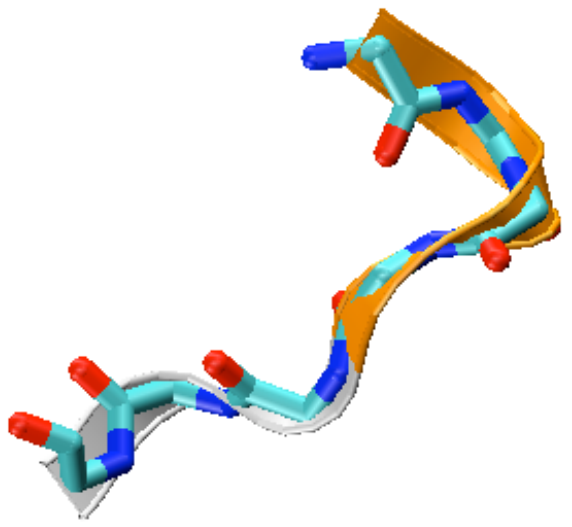 |
| β-Conformation<br><br>(s = 134 to 148) | 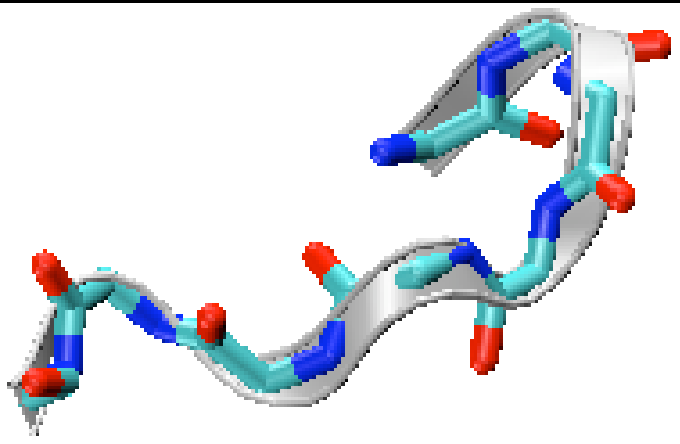 | **G32-Y37** loop region twists approx. at right angles at every sign change of τ like a 'staircase'.<br><br>(s = 122 to 144) | 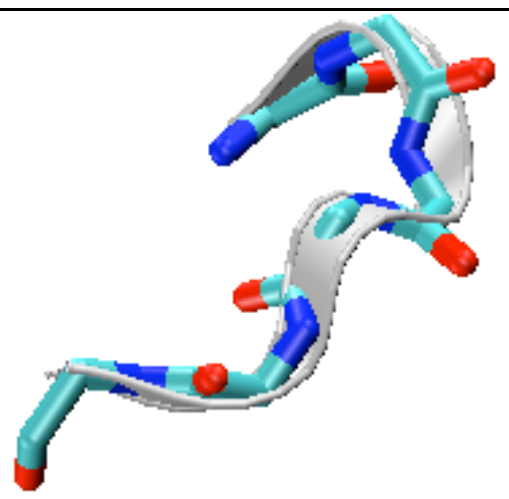 |
| β 3<br><br>(s = 160 to 174) | See ribbon diagram<br><br>Figure 3 | **C47-N57** folded beta strand<br><br>(s = 182 to 230) | See ribbon diagram<br><br>Figure 3 |
| Turn 4<br><br>(s = 174 to 186) | | **D59-D65** stretched right-handed loop; (s = 230 to 257) | |
| Helical segment<br><br>(s = 220 to 236) | | **D66-E71** distorted in 1UBQ, well-formed α-helix in 1A70.<br><br>(s = 257 to 280) | |
| Turn 4 - small β4 – Turn 5 - helical segment<br><br>(s = 174 to 236) | | Turn 4 – Helix 2 – Turn 5 - small β4 : rearranged relative to 1UBQ<br><br>(s = 230 to 300) | |
| Turn 6<br><br>(s = 244 to 258) | 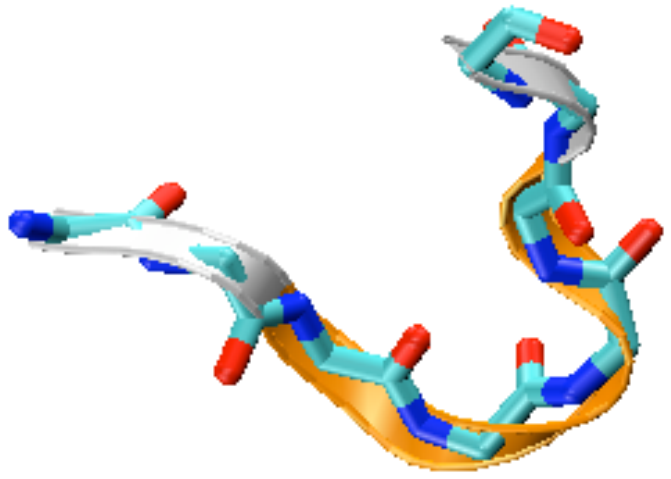 | **L75-A78** Partly helical as in 1UBQ. The first turn residue has positive τ pointing the rest of the turn downward (residues Q62 + K63 are oriented the same).<br><br>(s = 300 to 314) | 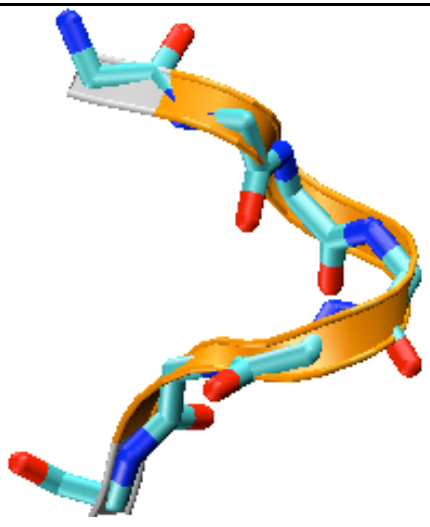 |
| β5<br><br>(s = 258 to 283) | Figure 3 | **A79-E88** long and folded like β3.<br><br>(s = 314 to 354) | Figure 3 |
| Turn 7<br><br>(s = 283 to end) | Figure 3 | **T89-A97** A turn within the β-strand in 1UBQ becomes a long helical loop; (s = 354 to end) | Figure 3 |

[a] Arc length values $s$ [Å$^{-1}$] are taken from Table 4 and Figure 2. The 3D structural units of 1UBQ and 1A70 have been prepared with VMD and oriented in such a way that the first two residues always point in the same direction for a pair. – Note that the Greek characters used in Tables 4 and 5 and in the text have been written out in Figures 2 and 3 to improve the readability of the latter.

**Table 6.** Grading system A-F used for the similarity test shown in Figure 5. [a]

| Grade | Same Length [a] of proteins | Different Length Property Criterion | Comment |
|---|---|---|---|
| A | 99% APSA agreement | Residues agree for P1-P8 | Similarity test is simplified by the agreement in length |
| B | ≥ 93% APSA agreement | Residues to be comparable agree for P1-P8 | In case of different protein length, residues without partners have to be eliminated |
| C | Not within the 20 domains | Comparable residues agree for P1-P6, differ for P7,P8 | Overall shape is maintained. 'Acceptable' differences are those where loop regions resemble distorted secondary structures. |
| D | No example | Comparable residues have similar P1-P3 and P5 | Different loop lengths may change sec. structure orientation. 'Acceptable' differences are discounted & κ(s) shows more similarity than τ(s) |
| E | No example | Comparable residues have similar P1-P2; P3 may differ | Differences in ordering of sec. structures cause different folds or at least topologies; some scattered turns contain resemblances |
| F | No example | Comparable residues have similar P1 | Different fold; some turns scattered in the domain are still similar with respect to shape and sign indicating similar supersec. structures |
| G | No example | P1-P8 are different | Different classes |

[a] Two proteins will be considered to be of the same length if the calculated arc lengths s agree within ± 5 Å. The following 8 structural properties P (ordered according to increasing detail) derived from the τ(s) diagrams of APSA are determined: P1) Ratio of helices and β-strands according to τ-patterns; P2) Number secondary structural units according to τ-patterns; P3) Order of secondary structural units according to τ-patterns; P4) Lengths of turns/loops connecting secondary structures according to arc length s; P5) Overall sign of torsion at the turns/loops according to τ(s) (τ is averaged over the whole turn by calculating area *under the curve* of the τ peaks); P6) τ-Sign of individual residues in the turns/loops; P7) Nature of the turns/loops whether the τ patterns resemble helical or extended conformations; 8) Match within assigned secondary structures with respect to distortions according to τ(s).

Figure 1

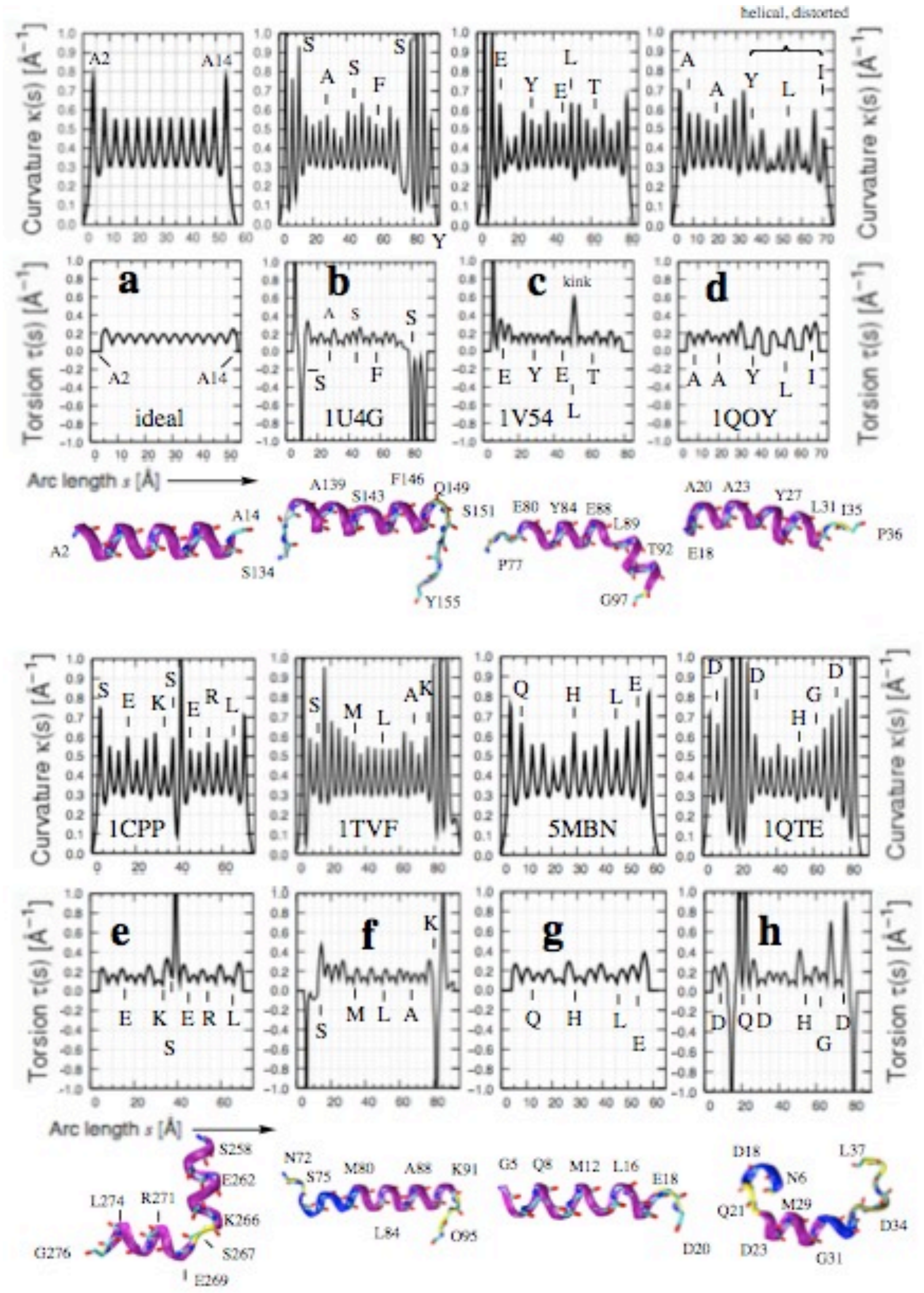

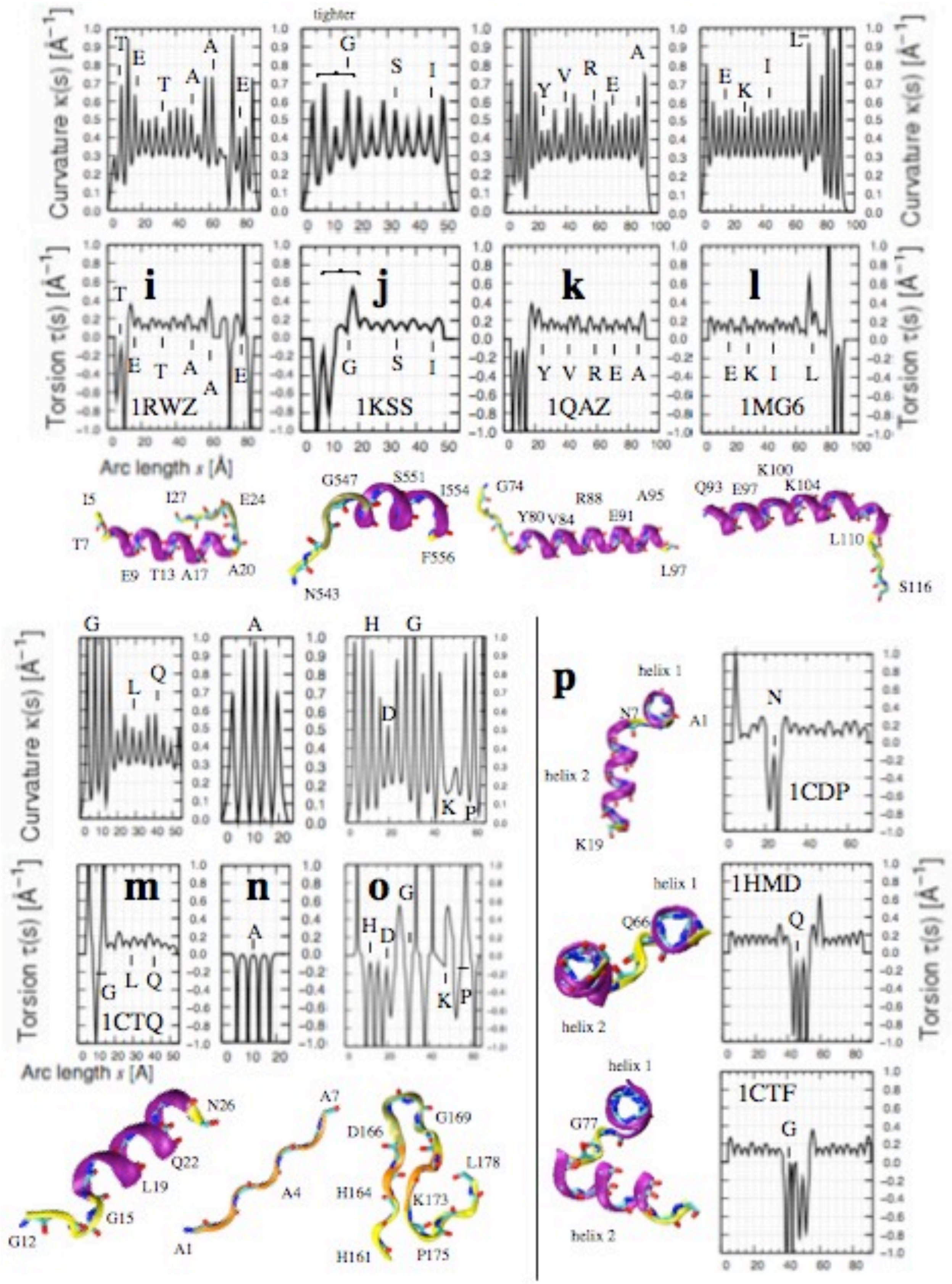

Figure 2a

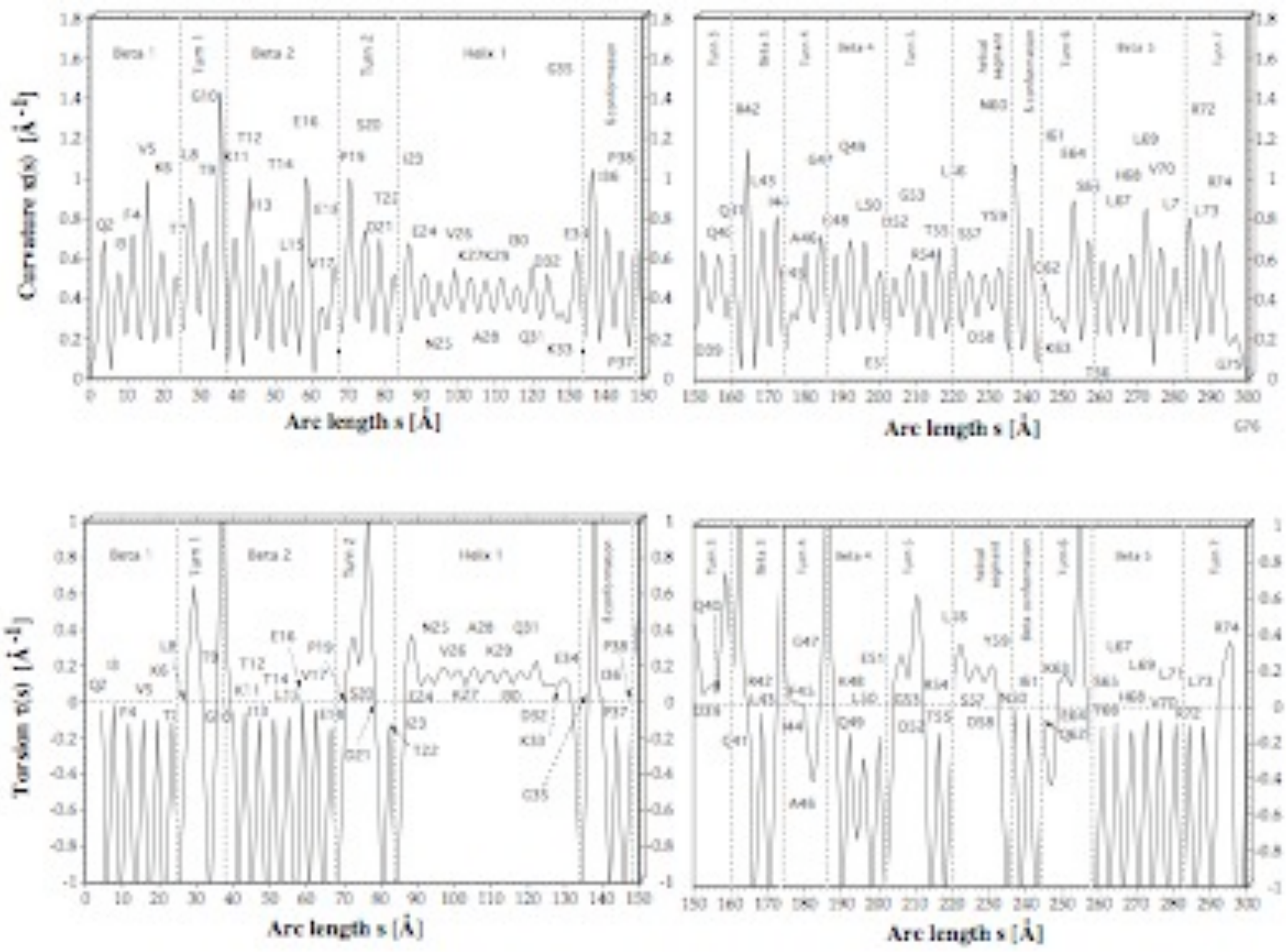

Figure 2b

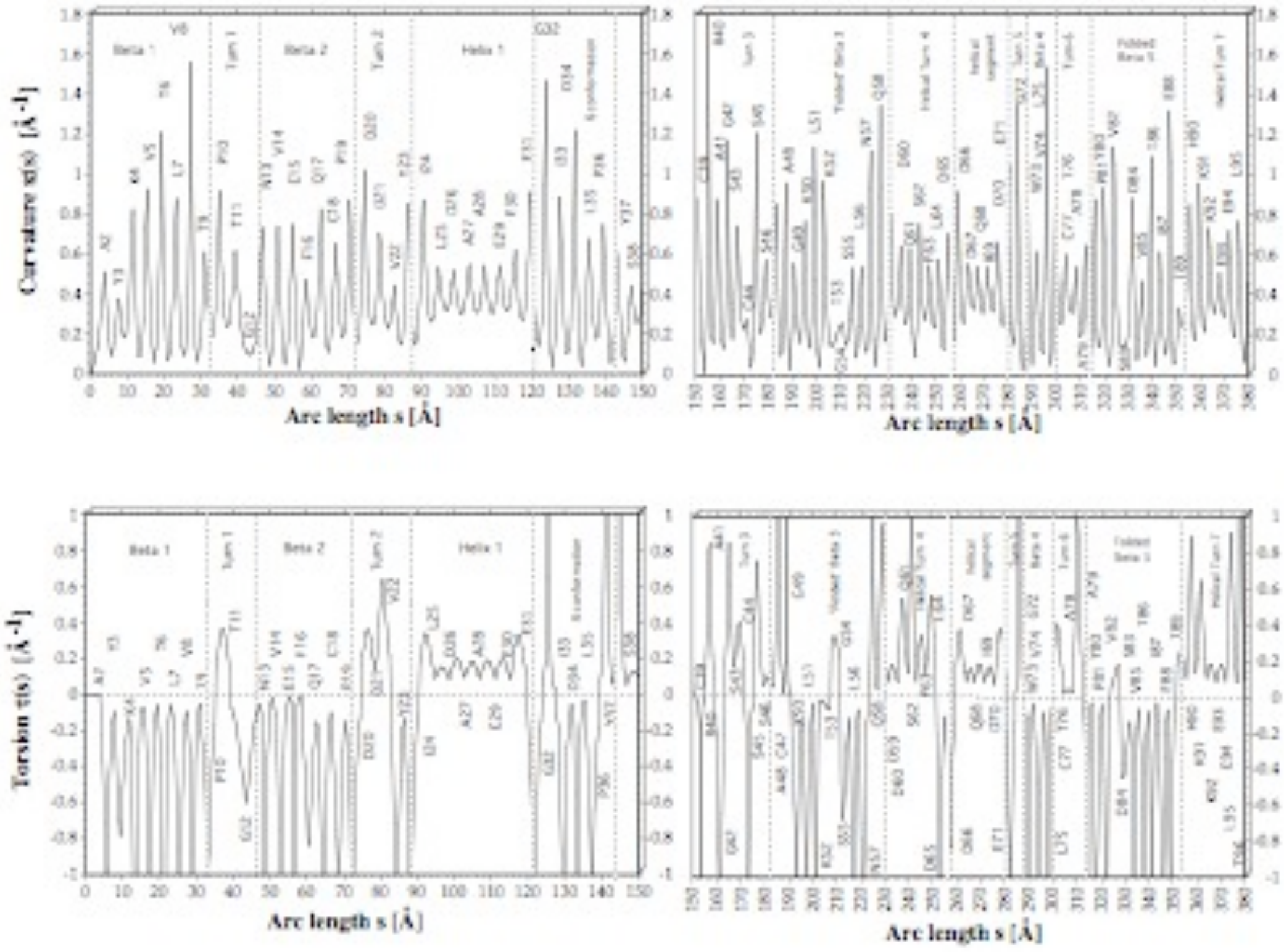

Figure 3

Ubiquitin (1UBQ) ribbon rendering

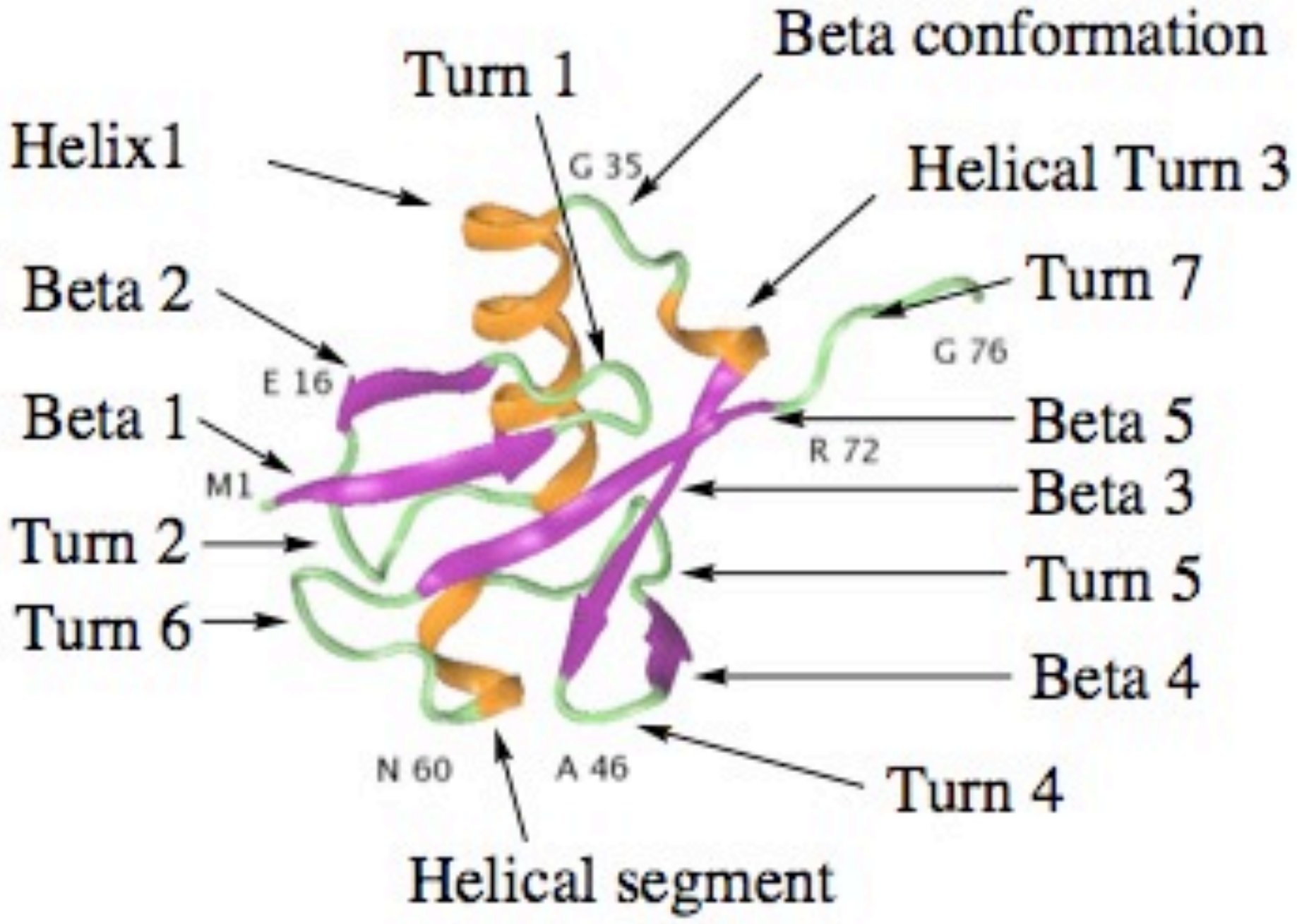

Ferridoxin (1A70) ribbon rendering

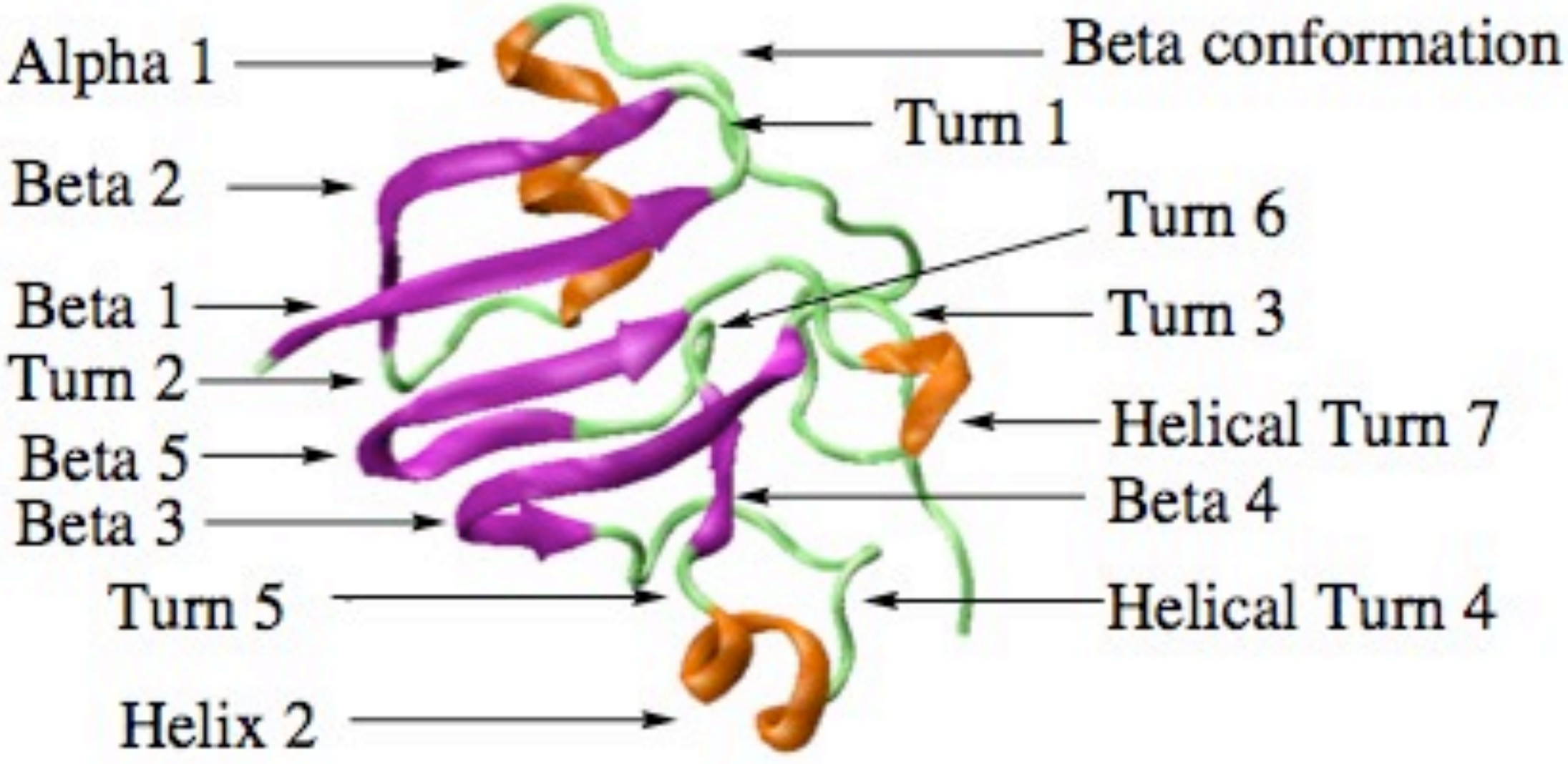

Figure 4

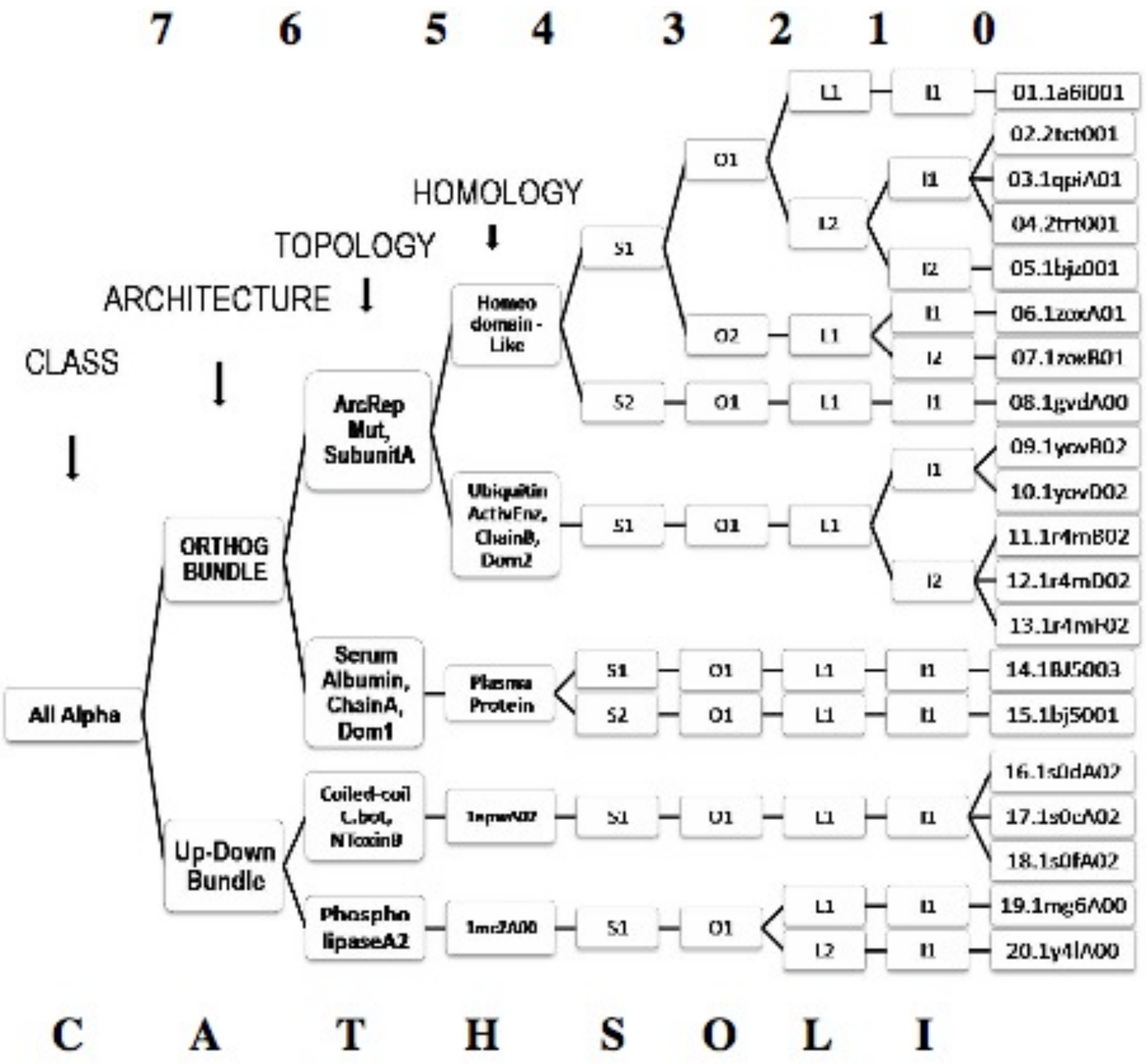
7 6 5 4 3 2 1 0
HOMOLOGY
TOPOLOGY
ARCHITECTURE
CLASS
All Alpha
ORTHOG BUNDLE
Up-Down Bundle
ArcRep Mut, SubunitA
Serum Albumin, ChainA, Dom1
Coiled-coil
Phospho lipaseA2
Homeo domain - Like
Plasma Protein
S1
S2
O1
O2
L1
L2
I1
I2
01.1a6I001
03.1qpiA01
05.1bjz001
06.1zoxA01
08.1gvdA00
10.1yovD02
11.1r4mB02
12.1r4mD02
15.1bj5001
17.1s0cA02
18.1s0fA02
19.1mg6A00
20.1y4lA00
C A T H S O L I

Figure 5

| | 1 | 2 | 3 | 4 | 5 | 6 | 7 | 8 | 9 | 10 | 11 | 12 | 13 | 14 | 15 | 16 | 17 | 18 | 19 | 20 |
|---|---|---|---|---|---|---|---|---|---|---|---|---|---|---|---|---|---|---|---|---|
| **1** | - | 2 | 2 | 2 | 2 | 3 | 3 | 4 | 5 | 5 | 5 | 5 | 5 | 6 | 6 | 7 | 7 | 7 | 7 | 7 |
| **2** | A | - | 0 | 0 | 1 | 3 | 3 | 4 | 5 | 5 | 5 | 5 | 5 | 6 | 6 | 7 | 7 | 7 | 7 | 7 |
| **3** | A | A | - | 0 | 1 | 3 | 3 | 4 | 5 | 5 | 5 | 5 | 5 | 6 | 6 | 7 | 7 | 7 | 7 | 7 |
| **4** | A | A | A | - | 1 | 3 | 3 | 4 | 5 | 5 | 5 | 5 | 5 | 6 | 6 | 7 | 7 | 7 | 7 | 7 |
| **5** | A | A | A | A | - | 3 | 3 | 4 | 5 | 5 | 5 | 5 | 5 | 6 | 6 | 7 | 7 | 7 | 7 | 7 |
| **6** | B | B | B | B | B | - | 1 | 4 | 5 | 5 | 5 | 5 | 5 | 6 | 6 | 7 | 7 | 7 | 7 | 7 |
| **7** | B | B | B | B | B | A | - | 4 | 5 | 5 | 5 | 5 | 5 | 6 | 6 | 7 | 7 | 7 | 7 | 7 |
| **8** | C | C | C | C | C | C | C | - | 5 | 5 | 5 | 5 | 5 | 6 | 6 | 7 | 7 | 7 | 7 | 7 |
| **9** | D | D | D | D | D | D | D | D | - | 0 | 1 | 1 | 1 | 6 | 6 | 7 | 7 | 7 | 7 | 7 |
| **10** | D | D | D | D | D | D | D | D | A | - | 1 | 1 | 1 | 6 | 6 | 7 | 7 | 7 | 7 | 7 |
| **11** | D | D | D | D | D | D | D | D | A | A | - | 0 | 0 | 6 | 6 | 7 | 7 | 7 | 7 | 7 |
| **12** | D | D | D | D | D | D | D | D | A | A | A | - | 0 | 6 | 6 | 7 | 7 | 7 | 7 | 7 |
| **13** | D | D | D | D | D | D | D | D | A | A | A | A | - | 6 | 6 | 7 | 7 | 7 | 7 | 7 |
| **14** | E | E | E | E | E | E* | E* | D | E | E | E* | E* | E* | - | 4 | 7 | 7 | 7 | 7 | 7 |
| **15** | E* | E* | E* | E* | E* | E | E | E | E | E | E | E | E | C | - | 7 | 7 | 7 | 7 | 7 |
| **16** | F | F | F | F | F | F | F | F | F | F | F | F | F | F | F | - | 0 | 0 | 6 | 6 |
| **17** | F | F | F | F | F | F | F | F | F | F | F | F | F | F | F | A | - | 0 | 6 | 6 |
| **18** | F | F | F | F | F | F | F | F | F | F | F | F | F | F | F | A | A | - | 6 | 6 |
| **19** | F | F | F | F | F | F | F | F | F | F | F | F | F | F | F | E | E | E | - | 2 |
| **20** | F | F | F | F | F | F | F | F | F | F | F | F | F | F | F | E | E | E | A | - |

Figure 6a

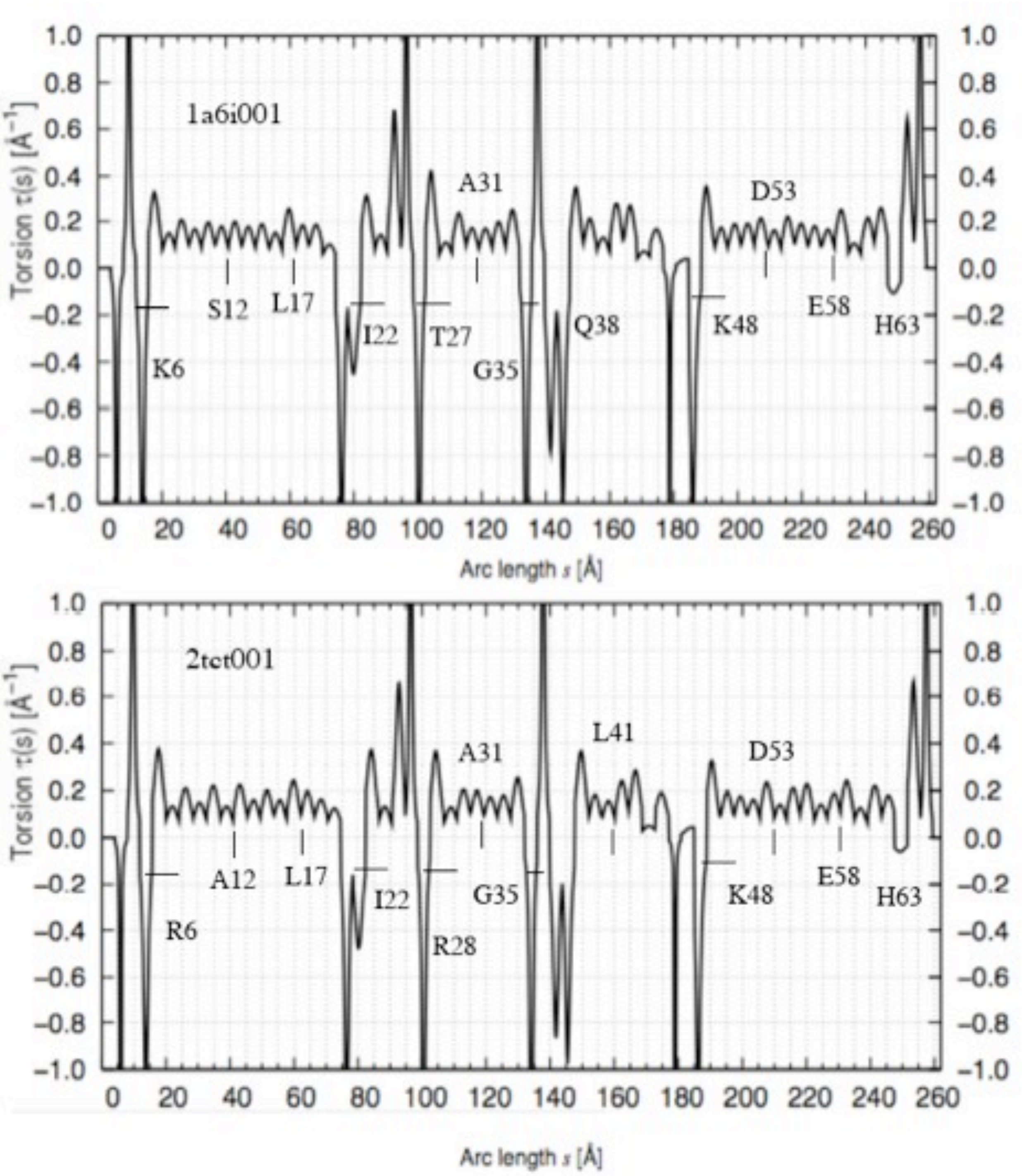

Figure 6b

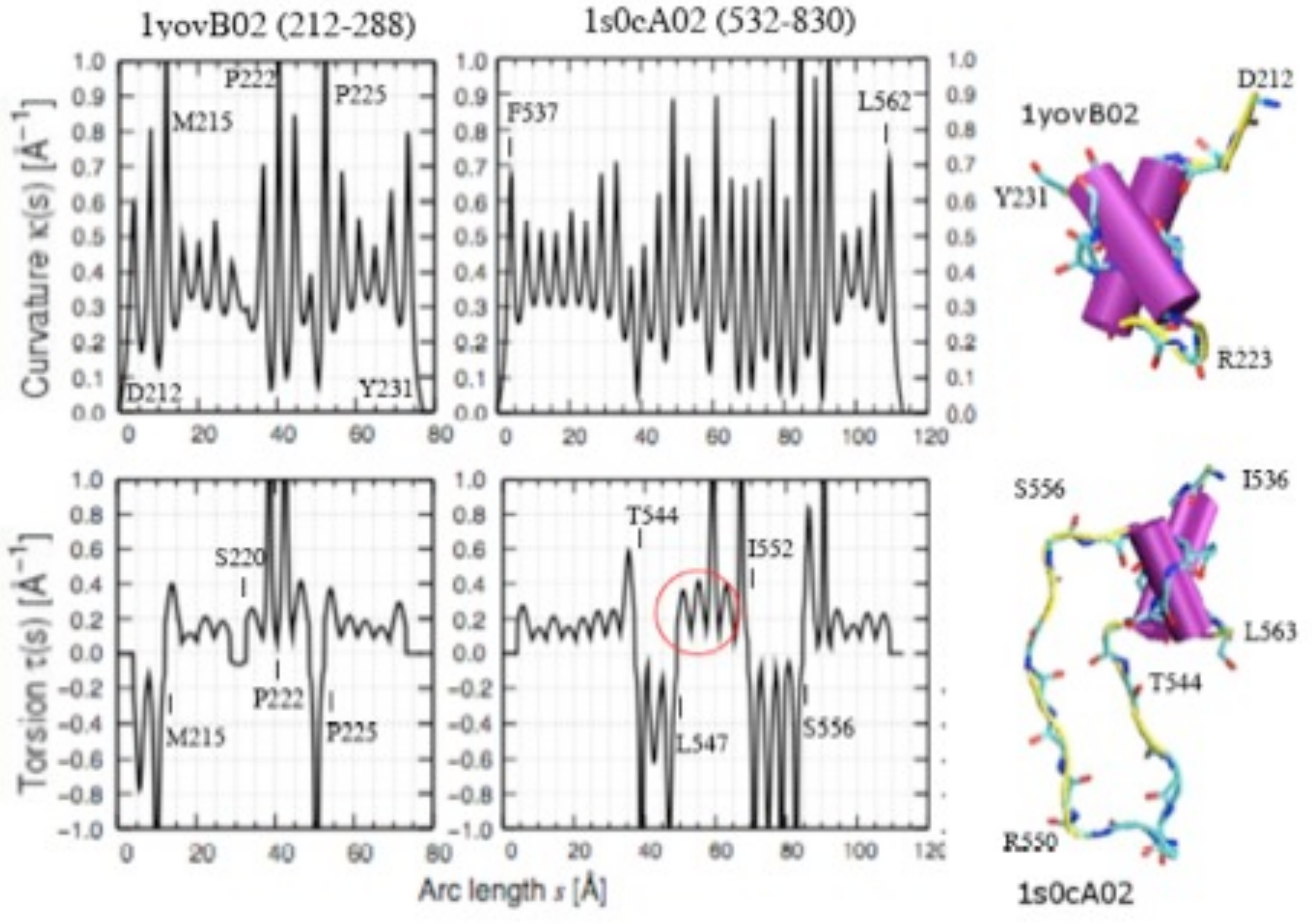

Scheme 1

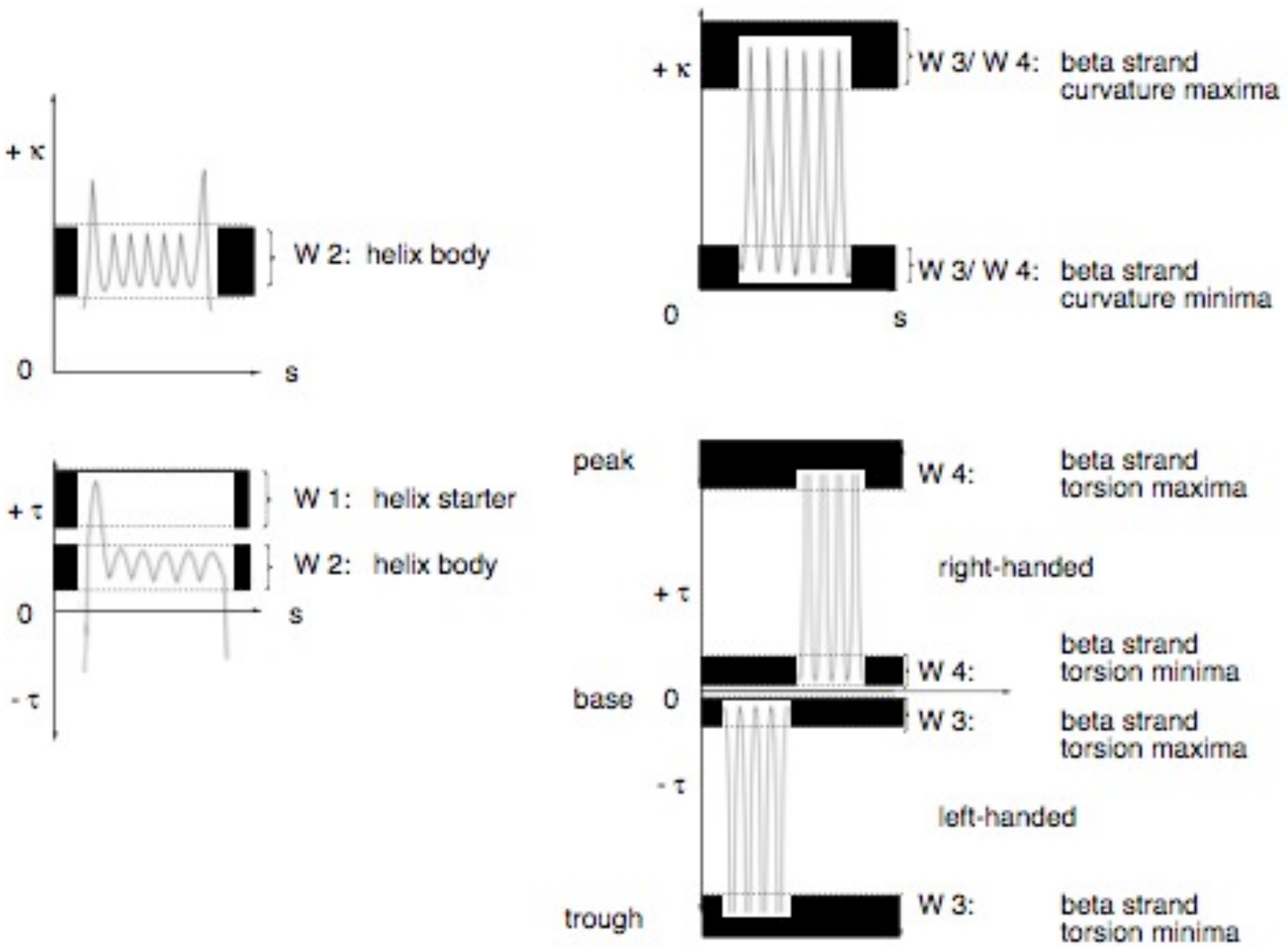

+ κ
W 2: helix body
0
s
+ τ
W 1: helix starter
W 2: helix body
0
s
- τ
+ κ
W 3/ W 4: beta strand curvature maxima
W 3/ W 4: beta strand curvature minima
0
s
peak
W 4:
beta strand torsion maxima
right-handed
+ τ
beta strand torsion minima
W 4:
base
0
W 3:
beta strand torsion maxima
- τ
left-handed
trough
W 3:
beta strand torsion minima

Scheme 2

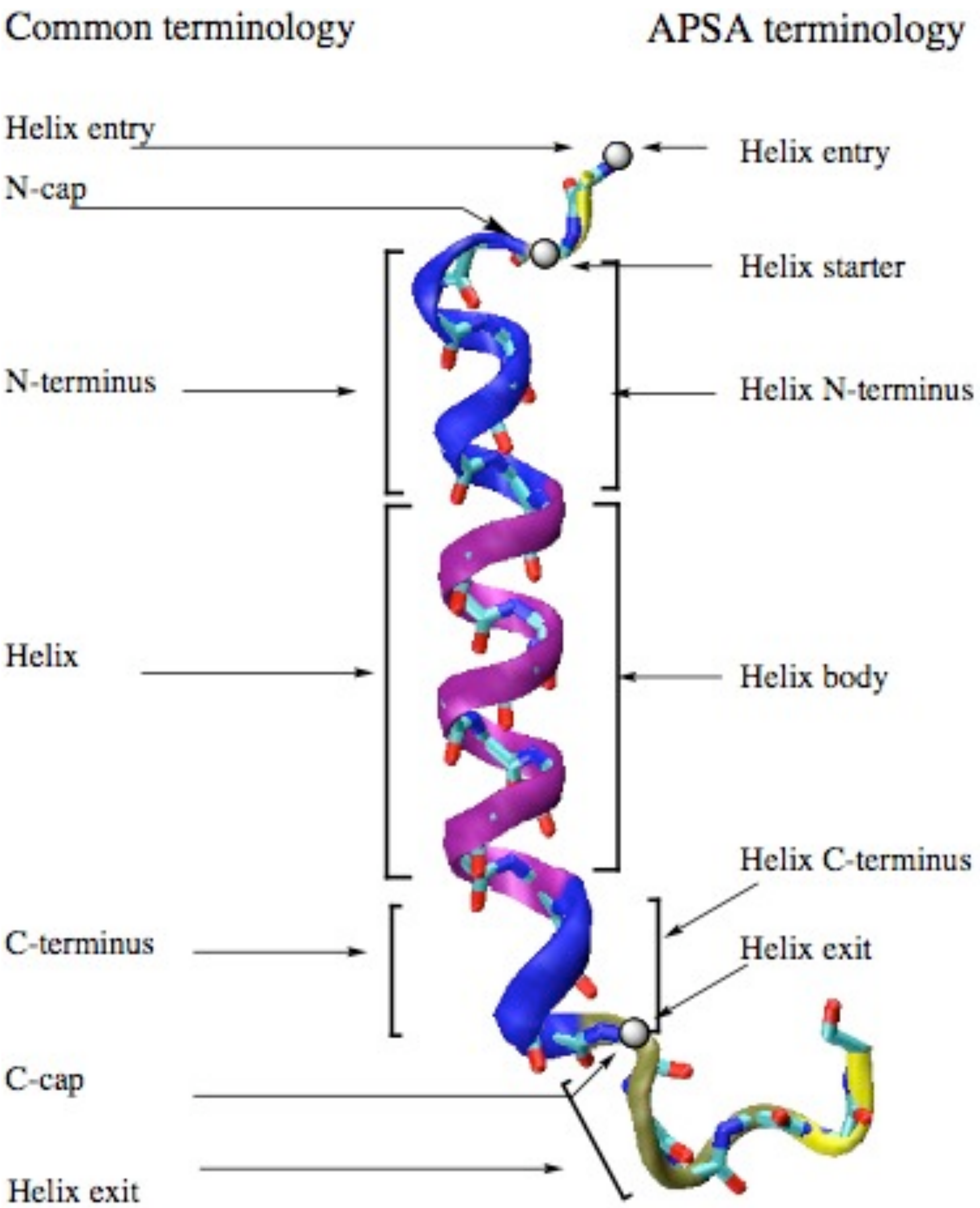